  \documentclass[journal,onecolumn,10pt]{IEEEtran}

\usepackage{latexsym}
\usepackage{amsmath,url}
\usepackage{amsfonts}
\usepackage{graphicx}
\usepackage{boxedminipage}
\usepackage{color} 
\usepackage{array}

\usepackage{ulem}
\usepackage{subfigure}

\usepackage{bm}        
\usepackage{latexsym}

\DeclareMathOperator*{\argmin}{arg\,min}  

\newcommand{\eqdef}{\stackrel{\textrm{\tiny def}}{=}}

\begin{document}

\title{Emitter Localisation  from Reception Timestamps \\ in Asynchronous  Networks} 

\author{\IEEEauthorblockN{Nicol\`{o}    Facchi, Francesco Gringoli, Fabio Ricciato, Andrea Toma}

\thanks{N. Facchi  and F. Gringoli are with the Department of Information Engineering, CNIT --- University of Brescia, Italy. Email n.facchi@unibs.it, francesco.gringoli@unibs.it.}
\thanks{F. Ricciato is with the Faculty of Computer and Information Science, University of Ljublijana, Slovenia. Email fabio.ricciato@fri.uni-lj.si.}
\thanks{A. Toma and F. Ricciato were with the Austrian Institute of Technology,  Austria, during the preparation of this work.}
}

%
%
\maketitle

\begin{abstract}
We address the problem of localising a mobile terminal (``blind" node)  in unknown position from a set of ``anchor" nodes  in known positions.   
The proposed method does not require any form of node synchronisation nor measurement (or control)  of the transmission times, which is difficult or anyway costly to achieve in practice.  It relies exclusively on reception timestamps collected by the anchor nodes, according to their local clocks,  that overhear packets transmitted by the blind node and by (at least one) other  transmitting node(s) in known position, e.g., other anchors.
The clock differences between the nodes  are not  eliminated  {\em ex ante}    through clock synchronisation, as in traditional ToA and TDoA methods. Instead, they are counteracted  {\em ex post}, during the data processing stage, leveraging the data redundancy that is intrinsic to the multiple reception of the same packet by different (anchor) nodes. 
We validate the proposed method in different experimental settings, indoor and outdoor, using exclusively low-cost Commercial-Off-The-Shelf WiFi devices, achieving sub-meter accuracy in full Line-of-Sight conditions and meter-level accuracy in mild NLOS environment. 
The proposed method does not require that the blind node participate actively to the localisation procedure and can use ``opportunistically" any legacy signal or packet available over-the-air for communication purposes.   
Considering the very minimal requirement on the system --- basically, only that  anchors in known positions are able to collect and share reception timestamps  --- the proposed approach can enable practical adoption of opportunistic and/or  cooperative localisation on top of existing radio communication systems.  
\end{abstract}


\section{Introduction}
Despite the widespread success of Global Navigation Satellite Systems (GNSS), alternative radio-based positioning systems remain of interest for those applications where GNSS signals are not available (e.g., indoor, tunnels, urban canyon) and/or the integration of GNSS receivers is not feasible due to size, energy or cost constraints.
In order to preserve cost-effectiveness and ease of adoption, it is desirable to develop  localisation methods that reuse {\em opportunistically} the legacy radio signals, protocols and  receivers which are anyway available for communication purposes.
We consider the problem of localising a transmitting node in unknown position, called ``blind node" hereafter,  based on the reception measurements collected by a set of nodes in known positions, called ``anchors".

One  possible approach is to rely on  {\em  power}  measurements, leveraging the Received Signal Strength Indicator (RSSI) available today in almost all Commercial Off-The-Shelf (COTS) transceivers to deploy RSSI-based localisation functions on top of existing wireless communication systems.
Both commercial and open-source tools are available for several technologies, including WiFi, Bluetooth, and IEEE 802.15.4 (e.g. \cite{awiloc, atheros2004,SN12}). 
Generally speaking, RSSI-based methods exhibit low accuracy, since in real environments the received power is  heavily affected by other factors than the distance from the transmitter, most prominently multipath propagation and antenna patterns. Several previous work seek to improve accuracy of RSSI-based methods using machine learning techniques that create and use detailed power loss maps for specific environments \cite{dina2009pam,ergen2014,kim13pervasive,fang2011}, in the direction of so called ``RSSI fingerprinting" techniques.

Another class of approaches is based on  {\em  time}  measurements. As signals travel at constant speed, the relation between reception time and distance is tighter than for received power, therefore time-based localisation is intrinsically superior to RSSI-based methods in terms of potential accuracy \cite{dardari2009}.  
However, the traditional approaches to time-based localisation require the system to support some form of {\em node synchronisation} and/or accurate {\em measurement or control of 
transmission times}. Both these functionalities are difficult or costly to achieve in practice \cite{sivrikaya04}. Node synchronisation requires the distribution of synchronisation signals, either via additional wired infrastructure or wireless protocols, and the implementation of dedicated functions into the devices which consume  additional  energy, bandwidth, and computation resources  \cite{sivrikaya04}. Controlling or measuring the actual transmission time is also difficult with standard hardware, due to unpredictable delay components between the construction of the packet  and the transfer to the network interface on one hand, and the intrinsic variability of channel access delays due, e.g., to MAC dynamics \cite{sivrikaya04}. For these reasons we are interested in time-based localisation methods that \emph{(i)} do not require any synchronisation across nodes and \emph{(ii)}  do not rely on  knowledge nor control of  \emph{transmission} times.

In standard Time-of-Arrival (ToA) methods  all nodes must use a common {synchronous} clock reference and must 
take  active part to the localisation procedure  \cite{dardari2009,alavi2006,alsindi2009}. The transmitting node attaches a transmission timestamp to the packet in order to allow the receiver to compute the travel time (also called ``time of flight") and from there derive the distance from the transmitter (one-way ranging).  
In two-way ranging methods instead,  node A sends a probe packet to node B  and the latter answers back with a reply packet that includes the information about the elapsed time between the reception of the probe and the transmission of the reply packet (also called ``turn-around time"). By subtracting this value from the elapsed time between the transmission of the probe packet and reception of the reply packet, node A is able to compute the distance (range) to B. Two-way ranging does not need clock synchronisation between the nodes, however it still requires accurate knowledge of (the difference between) pairs of {\em transmission and reception  times}. Moreover, it requires that the node to be localised (``blind node")  take active part into the ranging phase \cite{giustiniano2011}. 
With  Time-Difference of Arrival  (TDoA) methods an emitter node can be localised by a set of anchors in known positions that measure exclusively the reception times of the signals transmitted by the blind node \cite{yang2010}. This method relaxes the requirement on the transmission times,  but still needs tight synchronisation  between the anchor nodes for accurate localisation \cite{herath2013}.

Due to the requirements on node synchronisation and/or measurement or control of transmission time, it is impractical to deploy these methods on COTS devices \cite{krishan2009,gallo2013wmps,hoene2009,wang11}. To overcome these limitations, we consider here a localisation method that \emph{relies exclusively on reception timestamps and does not require any form of node synchronisation}. The proposed method can passively localise a (possibly unaware) blind node while it is executing standard communication operations, i.e., exchanging packets with a neighbouring node, without requiring its active participation to the localisation procedure. In other words, our method meets  all
the following system-level requirements:
\begin{itemize}
\item The blind node transmits beacons and/or packets (e.g. towards one or more other nodes, mobile or fixed)  but it does not take any active part to the localisation process. The localisation procedure involves exclusively a set of ``anchor" nodes in known positions. 
\item No synchronisation mechanism is in place between the anchor nodes nor between blind and anchors, i.e., we assume {\em asynchronous nodes}. 
\item Anchor nodes can measure only local {\em reception} times (i.e., timestamps) while the {\em transmission} times are unknown and can not be controlled, i.e., we assume {\em asynchronous signals}.
\end{itemize}
It can be easily seen that the above requirements  do not allow to adopt ToA, two-way ranging, TDoA nor other hybrid variants proposed recently in the literature 
(e.g. \cite{wang11,vaghefi13icassp,kim2014,Liu10,Zach14,Gholami13,Li06}) 
that anyway demand some combination of node synchronisation,  knowledge and/or control of transmission times and  blind node involvement. 

By fulfilling the above requirements,  the method described here can be adopted in a wide range of practical applications to  localise passively a mobile transmitter  based exclusively on the {\em local reception timestamps}  collected by other nodes placed in the surrounding environment. Waiving the requirements of node and signal synchronisation enables the implementation of localisation capabilities in legacy communication systems, which are typically completely asynchronous. Possible applications include indoor localisation of smartphones, customer tracking  for retail marketing intelligence \cite{tracking},  localisation in Wireless Sensor Networks (WSN), and extension (augmentation) of GNSS positioning in Cooperative Intelligent Transportation Systems (C-ITS) \cite{RicciatoCommag}.

While in traditional localisation systems clock differences are eliminated {\em ex-ante} by the synchronisation mechanism, we consider a model where clock differences are accounted for {\em ex-post}, i.e., they are either canceled or estimated during the data  processing stage, jointly with the estimation of the (unknown) blind position.
This approach is conceptually  similar  to the well-known ``double-difference" method used in GNSS to eliminate  residual errors on the carrier \emph{phase} for high-precision applications (see e.g. \cite{tiberius08}).   
Only a few recent pioneering works have started to consider this class of techniques for \emph{time} (instead of phase) measurements in the context of terrestrial (not GNSS) localisation systems \cite{maroti05,fan07,fan08,exel11,xu13,coluccia14,EUC14}.   
The method elaborated here belongs to the class of so-called {\em Differential} Time-Difference of Arrival (DTDoA). A detailed review of previous DTDoA literature is given later in the Related Work section. The main goal of our contribution is to demonstrate that DTDoA localisation  can deliver sub-meter accuracy (in Line-of-Sight conditions) using exclusively COTS receivers, without any additional clock synchronisation mechanism. 
To this end, we present for the first time experimental measurements obtained with a testbed setting consisting exclusively of COTS WiFi devices, in both indoor and outdoor  environments
\footnote{The experimental dataset used in this work along with the software developed to setup the testbed can be obtained from 
 \url{http://www.ing.unibs.it/~openfwwf/localisation}. }

The rest of this paper is organised as follows. In Section \ref{sec:methodology} we formalise the system model and describe step-by-step the proposed methodology in the basic scenario of receive-only anchors. 
In Section \ref{sec:experiments} we provide experimental results with COTS WiFi devices in two different testbed scenarios, indoor and outdoor, achieving meter-level accuracy.
In Section \ref{sec:active} we extend the method to work in scenarios with multiple active anchors that are able also to transmit (in addition to receive) packets: our experimental results  show that the positioning accuracy can be improved down to sub-meter level if all nodes are in Line-of-Sight (LOS).
In Section \ref{sec:activeNLOS} we extend our method to take advantage of frequency diversity via multi-channel measurements, in this way we can maintain meter-level accuracy  also in mild Non-Line-of-Sight (NLOS) conditions.
In Section \ref{sec:discussion} we discuss future research directions for progressing and extending the work. 
In Section \ref{sec:related} we position our method with respect to existing literature. 
Finally, we conclude in Section \ref{sec:conclusions}.

\begin{figure}
\centering
\subfigure[]{\includegraphics[width=8cm]{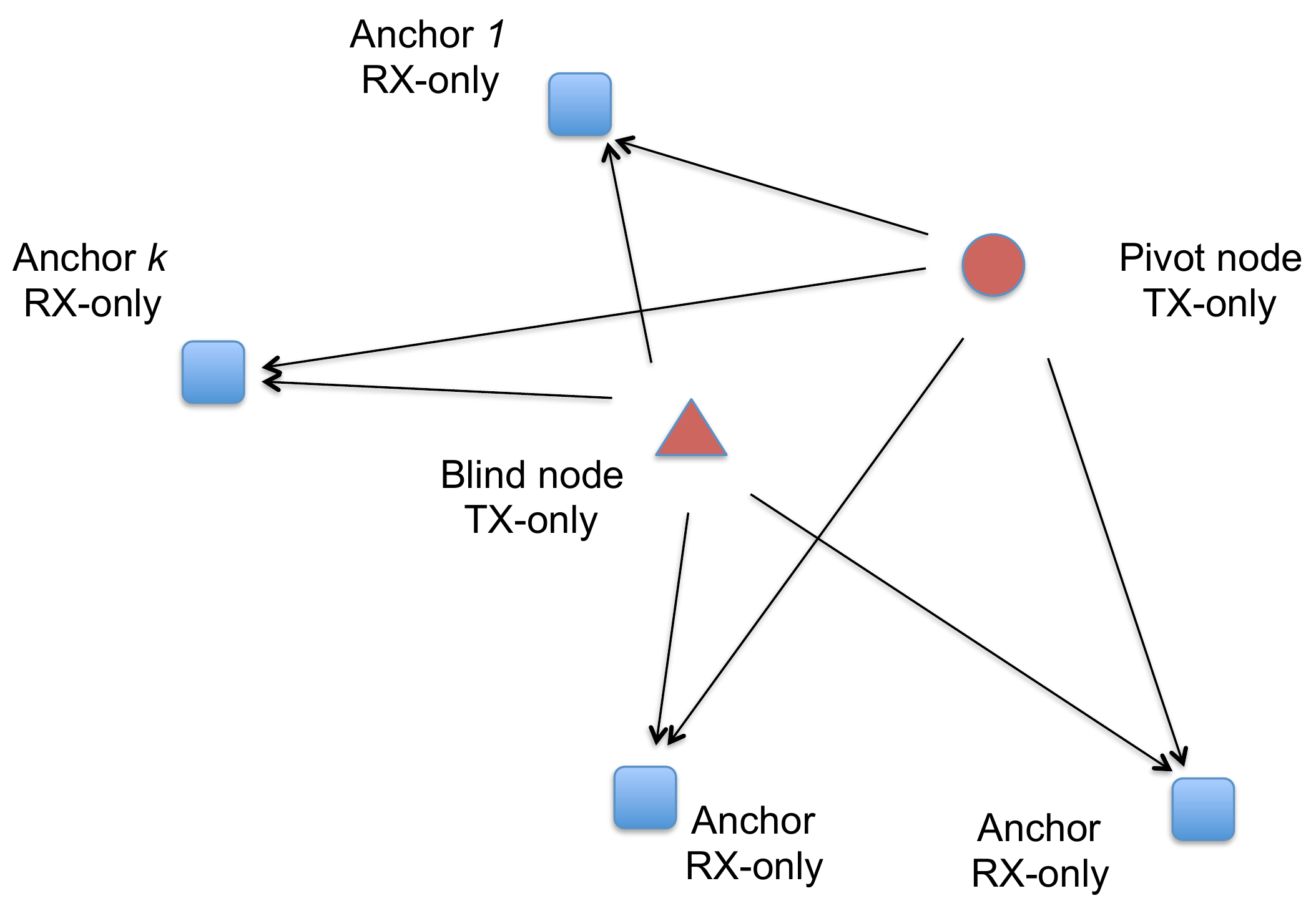}
\label{fig:scenario1}} \; \; \; \;
\subfigure[]{\includegraphics[width=7cm]{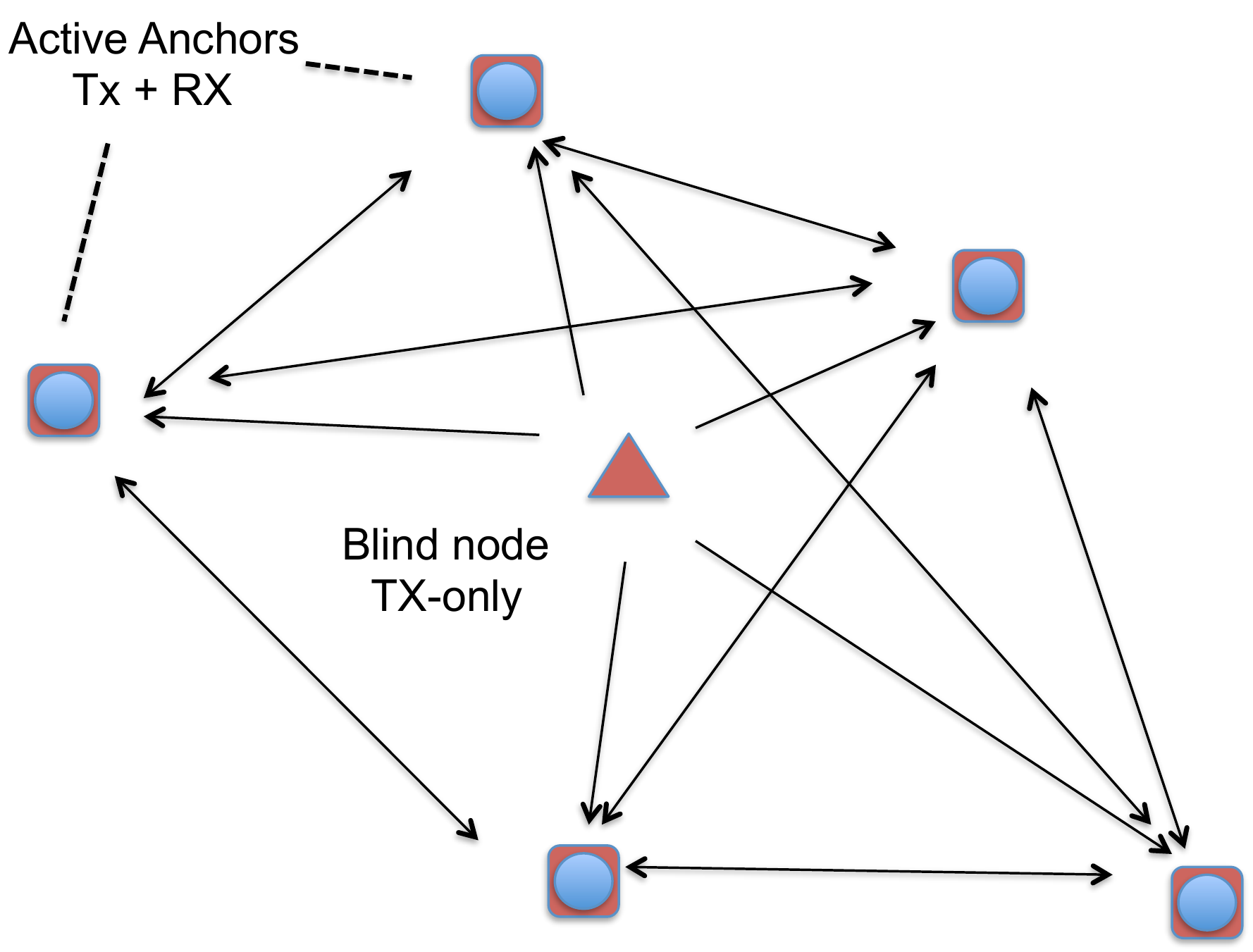}
\label{fig:scen2}}
\caption{Reference scenarios. (a) Basic scenario with multiple receive-only anchors and a single pivot node. (b) Extended scenario with multiple active anchors.}
\label{fig:scenboth}
\end{figure}

\section{Methodology}\label{sec:methodology}

\subsection{System model}\label{sec:scenario}
For the sake of simplicity we consider a 2D localisation problem, being the extension to 3D localisation quite straightforward.

The basic reference scenario represented in Fig. \ref{fig:scenario1} involves three types of radio nodes: the {\bf blind node}, one {\bf pivot node} and a set of (minimum three) {\bf anchors}. 
The positions of the pivot node and  of all anchor nodes are known, and the goal is to estimate the (unknown) position of the blind node. 
We consider a basic scenario where the blind and pivot nodes emit packets that can be overheard and timestamped by the anchor nodes. 
In other words, for the purpose of collecting timing measurements,  the anchors act as passive receivers (Rx-only) while the blind and pivot act as pure emitters (Tx-only).
No synchronisation method is adopted, i.e., the system is completely asynchronous. 
In a nutshell, the role of the pivot node is to provide reference signals (packets),  transmitted from a known location,  that compensate the lack of synchronisation between the anchors. 
In practical setting, the blind and pivot nodes could map,  respectively,  to a mobile terminal (e.g. smartphone) and to a fixed Access Point (e.g. WiFi or Bluetooth). These two nodes normally  exchange data and acknowledgment packets between themselves, however they are not required to take  active part in the localisation procedure, i.e., they do not need to perform any additional function besides the standard communication protocol operations. 
On the anchor side, the only requirement is that they timestamp every received packet and communicate the  collected timestamps and the associated packet identifiers to a central computation server in charge of  running the localisation algorithm, taking in input the (known) positions of the anchor and of the pivot. 

For the sake of simplicity we consider here a centralised scheme and a fully trusted environment, where measurement data and computation are concentrated in a single central server, and all the input data are correct (up to unintentional measurement errors) and shared openly. However, considering that the proposed algorithm consists of a series of simple linear operations, it should be possible to develop more sophisticated variants amenable to distributed computation, possibly taking into account additional requirements for preservation of sensitive data (e.g. anchor node location) and/or robustness against malicious users. These aspects fall outside the focus of the present work and are left as prominent directions for future research.

\subsection{Notation and basic equations}
With reference to the basic scenario depicted in Fig. \ref{fig:scenario1} we introduce the following notation.
We shall use the symbol ``$*$" and  ``$o$" to label the variables (e.g. transmission and reception times) associated to packets transmitted  by the blind and pivot node, respectively. 
Anchor nodes are indexed in $k$, with $k=1,\ldots, K$,  and $K \geq 3 $ denotes the total number of available anchors.
We define the following variables:
\begin{itemize}
\item $t_{i}^*$ [resp.  $t_{i}^o$] denotes the \emph{transmission time} (according to an ideal absolute clock) of the $i$th packet transmitted by the blind node [resp. pivot].
\item $r_{i,k}^*$ [resp.  $r_{i,k}^o$] denotes the \emph{reception  time} (according to an ideal absolute clock) at anchor node $k$ of the $i$th packet transmitted by the blind node [resp. pivot].
\item $s_{i,k}^*$ [resp.  $s_{i,k}^o$] denotes the \emph{reception  timestamp}  recorded at anchor node $k$ according  to its local clock for the $i$th packet transmitted by the blind node [resp. pivot].
\item $\bm{p}$,  $\bm{p}^o$ and $\bm{p}_k$  denote the position vectors,  respectively  of the blind, pivot and generic $k$th anchor. 
\item $d_{k} \eqdef \|\bm{p} - \bm{p}_k \| $  and  $d_{k}^o \eqdef \|\bm{p}^o - \bm{p}_k \| $  denote the euclidean distances between anchor $k$ and,  respectively, the blind node and pivot node.
\end{itemize}

\begin{figure}[t]
\centering
\includegraphics[width=9cm]{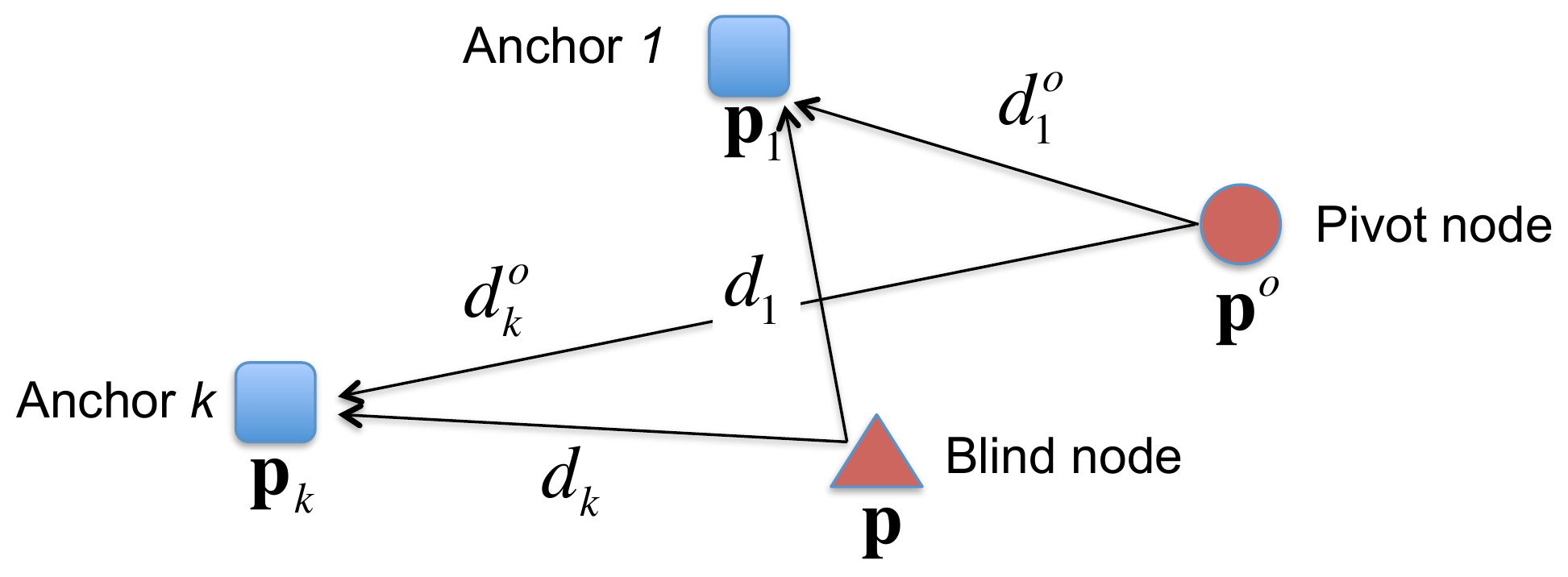}
\caption{Example of node quadruplet. $K-1$ total quadruplets are formed by considering different anchors $k \in \{2, \ldots, K\}$.}\label{fig:toy4plet}
\end{figure}

The proposed method is based on the estimation of  range differences among subsets of four nodes --- hereafter referred to as ``quadruplets" ---  consisting of a pair of transmitters, namely the blind node and the pivot, and a pair of receiving anchors. A generic quadruplet is depicted in Fig. \ref{fig:toy4plet}. One of the $K$ anchor nodes shall serve as main reference:  without loss of generality we pick the first anchor, with index $k=1$, as the reference anchor. Since all quadruplets have three nodes in common --- blind, pivot and reference anchor --- and differ only for the remaining anchor, we obtain $K-1$ quadruplets. 
For each anchor (equivalently: quadruplet) $k \in \{2, \ldots, K\}$ we  define the following {\em differential ranges}:
\begin{subequations}
\begin{equation}  \Delta_k \eqdef d_{k}  - d_{1}  = \|\bm{p} - \bm{p}_k \|  - \|\bm{p} - \bm{p}_1 \|,    \label{eq:defDelta1} \end{equation}
\begin{equation}  \Delta_k^o \eqdef d_{k}^o  - d_{1}^o  =  \|\bm{p}^o - \bm{p}_k \|  - \|\bm{p}^o - \bm{p}_1 \|.   \label{eq:defDelta2} \end{equation}
\label{eq:defDelta0} 
\end{subequations}
Note that $\Delta_k^o$ is fully known, while $\Delta_k$ is unknown as it contains the (unknown) blind node position   $\bm{p}$. 
Considering two packets $i$ and $j$ transmitted, respectively, by the blind  and  pivot node, 
their  transmission and reception times are linked by the following equations ($k = 1, \ldots, K$):
\begin{subequations}
  \begin{equation}    r_{i,k}^* = t_{i}^* + \frac{d_{k}}{c}  \label{eq:link1}\end{equation}
  \begin{equation}    r_{j,k}^o = t_{j}^o + \frac{d_{k}^o}{c}.  \label{eq:link2}\end{equation}
  \label{eq:12}
\end{subequations}
The relation between the  reception time $r_{i,k}$  and the associated timestamp $s_{i,k}$ can be
  modelled as follows: 
 \begin{subequations}
  \begin{equation}   s_{i,k}^*  = \alpha_k + \left( 1 + \beta_k \right)  r_{i,k}^*  +        h_k \left(  r_{i,k}^* \right) + \omega_{i,k}^* 
 \label{eq:link1b}\end{equation}
  \begin{equation}   s_{j,k}^o  = \alpha_k + \left( 1 + \beta_k \right)  r_{j,k}^o  + h_k \left(  r_{j,k}^o \right) + \omega_{j,k}^o 
  \label{eq:link2b}\end{equation}
  \label{eq:12b}
\end{subequations}
wherein
\begin{itemize}
\item $ \alpha_k $ denotes the (unknown) initial temporal offset of node $k$ clock with respect to the  ideal time reference.
\item $ \beta_k $ denotes the (unknown) value of the 	\emph{relative} frequency deviation  of node $k$ clock oscillator with respect to its nominal  frequency. It is an adimensional quantity that can take positive or negative values, normally limited to a few ppm.  For a generic anchor, the value of $ \beta_k $  can be considered constant at the  time-scales of interest for our problem (a few seconds or fractions thereof). Therefore, we obtain a  trend component in the clock error that is linear in time and varies from anchor to anchor.
\item $h_k \left(  \tau \right) $ represents a  {\em slowly-varying} function of time $\tau$ that captures any non-stationary component of clock error (e.g.  fluctuations of frequency offset due to temperature variations). 
\item  $\omega_{i,k}^*$ and  $\omega_{j,k}^o$   represent the residual {\em fast-varying} error components. They absorb all  non-systematic error components --- e.g., random delays in the receiver processing chain, timestamp quantization or truncation error, etc. --- and can be modelled as independent random variables with general distribution, not necessarily zero-mean.  
\end{itemize}

\subsection{Algorithm overview}

\begin{figure}[t]
\centering
\includegraphics[width=10cm]{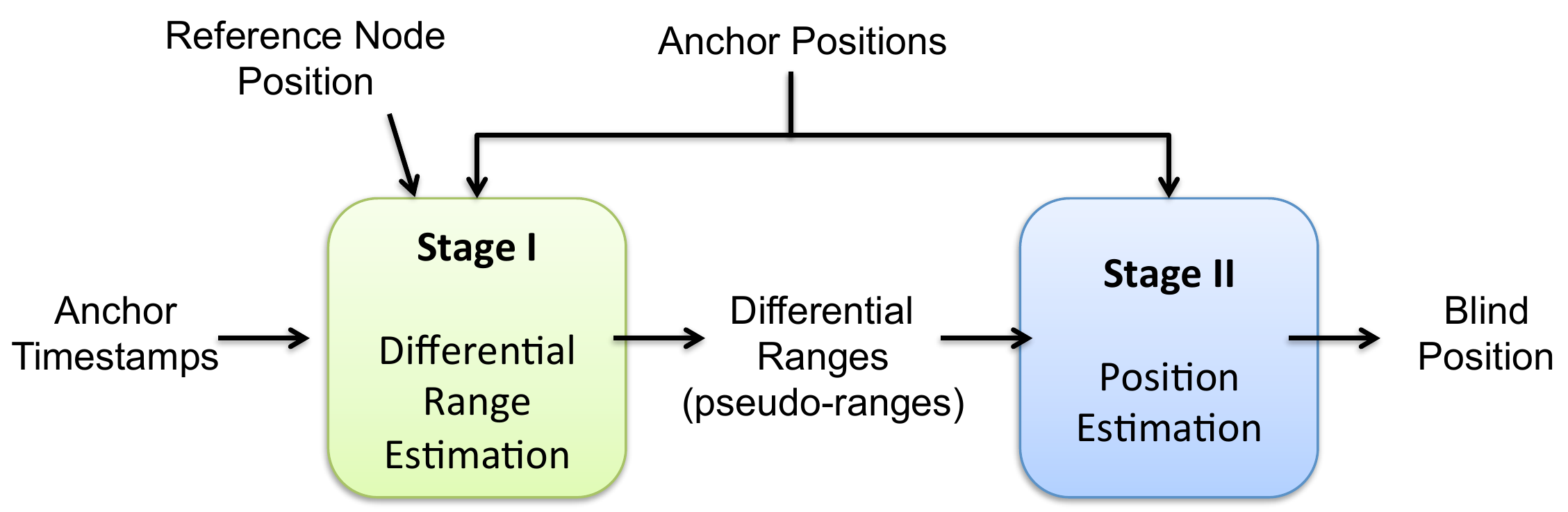}
\caption{Flow diagram of the proposed localisation procedure.}\label{fig:scheme1}
\end{figure}

In the reference scenario outlined above, the only measurable quantities are the receiver timestamps collected by anchors. This is the input to the algorithm, along with the positions of all anchors and pivot nodes, i.e., $\{\bm{p}^o,  \bm{p}_k, \; k=1, \ldots, K \}$. The desired output is (an estimate of) the blind position $\bm{p}$. 
The localisation procedure is split into two stages, as depicted in Fig. \ref{fig:scheme1}, with  a set of (differential) range measurements playing the role of intermediate variables. 
In the first stage, a set of differential ranges $\left\{ \Delta_{k}^*, \; k=2, \dots, K \right\}$ --- one for every quadruplet --- is computed based on the input timestamps and node locations.
In the second stage, the  set of differential ranges is used to compute the final blind position.

In the first (differential) ranging phase we must cope with two types of unknown quantities: the blind node position, and various error components for each anchor clock.  In order to produce differential ranges with acceptable accuracy we need to eliminate, or at least reduce, the systematic clock error components embedded in the raw timestamps. In our algorithm this is achieved through subtraction  and rescaling of timestamps, as explained hereafter.

\subsection{Estimation of   inter-anchor clock frequency ratio}\label{sec:gamma}
In the first step we run a simple linear regression over a block of $M$  data points  $ \left\{s_{j,k}^o, s_{j,1}^o \right\}_{j=1 \dots M}$, i.e., pairs of reception timestamps collected at  anchor $k$ and at the reference anchor for a block of $M$ consecutive packets transmitted by the pivot node ---  note that using the packets transmitted by the blind node $ \left\{s_{i,k}^*, s_{i,1}^* \right\}_i $   would be just equivalent.  The slope factor of the regressed line, hereafter denoted by $\hat{\gamma}_k$, provides an estimate of the ratio between the {\em actual} clock frequencies of the two anchors, i.e. $ \hat{\gamma}_k   \approx  \gamma_k  \eqdef  \frac{1+\beta_1}{1+ \beta_k} $ for  $k =2 \dots K$.    
This value will be used later to correct the timestamp measurements collected at anchor $k$ before subtraction of the corresponding timestamps at the reference anchor, as explained below. 

\subsection{Timestamp Subtraction}\label{sec:diff}

Consider a pair of neighbouring packets $i$ and $j$ transmitted respectively by the blind node and by the pivot node in close by instants, e.g., one data packet and the corresponding acknowledgment. For these two packets, we consider a pair of receiving anchors: the reference anchor (for which  $k=1$) and another generic anchor $k \in \{2, \ldots, K \}$, as sketched in Fig. \ref{fig:toy4plet}.
Recalling \eqref{eq:12} and \eqref{eq:12b} we can write:
  \begin{equation}   
  s_{j,1}^o - s_{i,1}^*   =  \left( 1 + \beta_1 \right) \cdot \left(   r_{j,1}^o - r_{i,1}^*  \right) + n_{ij,1} 
\label{eq:diff1}
 \end{equation}
wherein  for a generic anchor $k=1 \dots K$ (including the reference anchor)  the error term
\[  
n_{ij,k} \eqdef    \omega_{j,k}^o  -  \omega_{i,k}^* + h_k \left(  r_{j,k}^o \right) - h_k \left(  r_{i,k}^* \right) 
\]
cumulates the two individual random terms  and the residual difference between the non-stationary error components.
The latter is small, due to the fact that the reception times are close by and the function $h_k(\tau)$ is slowly varying, formally:   $r_{j,k}^o \approx r_{i,k}^*  \Rightarrow    h_k \left(  r_{j,k}^o \right) - h_k \left(  r_{i,k}^* \right) \approx 0 $. 
By symmetry, the term $n_{ij,k}$ can be  modelled as a zero-mean random variable even if the individual components   $\omega_{j,k}^o$ and $ \omega_{i,k}^* $ have non-zero mean, since by definition they are (independent) variables extracted by the same process at anchor $k$. In other words, $n_{ij,k}$ represents a measurement noise term. 

For the same packet pair $(i,j)$ we obtain an  equation similar to \eqref{eq:diff1} for every other anchor $k \geq 2$: 
  \begin{equation*}   
  s_{j,k}^o - s_{i,k}^*   =  \left( 1 + \beta_k \right) \cdot \left(   r_{j,k}^o - r_{i,k}^*  \right) + n_{ij,k} 
\label{eq:diffk}
 \end{equation*}
and by multiplying by $\gamma_k $ we obtain:
  \begin{equation}   
  \gamma_k \cdot \left( s_{j,k}^o - s_{i,k}^* \right)   =   \left( 1 + \beta_1 \right) \cdot \left(   r_{j,k}^o - r_{i,k}^*  \right) + \gamma_k \cdot n_{ij,k}
\label{eq:diffkgamma}
 \end{equation}
(note the change from $\beta_k$  to $\beta_1$ in the  right-hand term). 
Taking the difference of \eqref{eq:diff1}  and \eqref{eq:diffkgamma} and recalling \eqref{eq:defDelta0} and \eqref{eq:12}  we obtain:
  \begin{equation}   
S_{ij,k}   =  \left( 1 + \beta_1 \right) \cdot \frac{  \Delta_k  - \Delta_k^o  }{c}  + n'_{ij,k} 
\label{eq:finaldiff}
 \end{equation}
wherein 
\begin{equation}   
S_{ij,k} \eqdef      s_{j,1}^o - s_{i,1}^*  - \gamma_k \cdot \left( s_{j,k}^o - s_{i,k}^* \right) \label{eq:s}
\end{equation}
is the compound double difference of timestamps adjusted for  {\em relative} clock skew  (by $\gamma_k$ rescaling), and 
\begin{equation*}   
n'_{ij,k} \eqdef      n_{ij,1} - \gamma_k \cdot n_{ij,k}
\label{eq:N}
\end{equation*}
is a compound noise term.
By using the estimate  $\hat{\gamma}_k$ (obtained previously from linear regression) in place of $\gamma_k$ in \eqref{eq:s} we compute an approximation  $\hat{S}_{ij,k}$ of $S_{ij,k}$, 
 formally:
\begin{equation}   
\hat{S}_{ij,k} \eqdef      s_{j,1}^o - s_{i,1}^*  - \hat{\gamma}_k \cdot \left( s_{j,k}^o - s_{i,k}^* \right). \label{eq:shat}
\end{equation}
The approximation error on ${S}_{ij,k} $
 is also zero-mean and  can be absorbed into a total noise term, i.e.,  
$N_{ij,k} \eqdef n'_{ij,k} +   \hat{S}_{ij,k}  - {S}_{ij,k} $.
 With these positions we finally obtain:
  \begin{equation}   
\hat{S}_{ij,k}   =  \left( 1 + \beta_1 \right) \cdot  \frac{    \Delta_k  - \Delta_k^o  }{c}  + N_{ij,k} 
\label{eq:main2}
\end{equation}

\begin{figure}
\centering
\includegraphics[width=10cm]{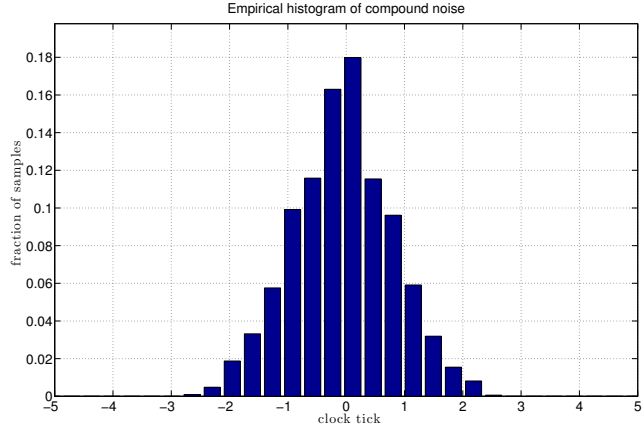}
\caption{Empirical distribution of compound measurement noise $N_{ij,k}$ for one sample experiment.}
\label{fig:Sijk}
\end{figure}

Fig.  \ref{fig:Sijk} reports the empirical distribution of the compound noise term $N_{ij,k}$ expressed in ticks of 22 MHz clock. Each tick corresponds in space to $\frac{22 \cdot 10^6 }{3 \cdot 10^8} =13.6$ meters.
The measurements  were obtained from \eqref{eq:main2}  in a symmetric squared topology, with the four nodes arranged in a way to ensure $\Delta_k  = \Delta_k^o \Rightarrow \hat{S}_{ij,k}  = N_{ij,k} $.  It can be seen that  the compound noise term has a zero-mean bell-like distribution.
Although the initial measurement error on each individual data point may seem quite large, up to 30 meters, the composition of multiple data points in time (by averaging multiple measurements for the same quadruplet) and space (by fusing measurements from different quadruplets) will drive the final position error down to meter level.

\subsection{Averaging and differential range computation}

In our experiments the packets transmitted by the  blind and pivot nodes alternate, so that at every receiver each packet from the blind node can be univocally associated to the successive packet from the pivot. Therefore, we can omit the index $j$ and label exclusively by $i$ (instead of $ij$) the individual measurement originated by the generic packet pair. Furthermore, we set $\mu \eqdef \left( 1+ \beta_1\right)^{-1}$. With these notational simplifications \eqref{eq:main2} rewrites as: 
\begin{equation}   
\mu \cdot \hat{S}_{i,k}   =   \frac{    \Delta_k  - \Delta_k^o  }{c}  + N_{i,k}.
\label{eq:main2simple}
\end{equation}
wherein the scalar term $\mu \approx 1$ accounts for the (unknown) frequency offset between the reference anchor and an absolute ideal clock.
Recall that rescaling by $\hat{\gamma}_k$ in \eqref{eq:shat} has compensated for the {\em differences} in  frequency offset between the anchor nodes,
thus aligning the whole system to a single common clock frequency (specifically the one of the reference anchor)  but has not eliminated the {\em absolute} frequency offset of the system with respect to an ideal absolute  clock. In other words, the effect of correcting for the \emph{relative} frequency offset between the anchors  emulates a system running with a \emph{single common clock} affected by an unknown (small, but non-zero) skew term captured by $\mu$. As we will show later, the position estimation process is practically insensitive to a small common skew term, meaning that the approximation  $\mu = 1$ will introduce a negligible error on the final position (see proof in Appendix).


In order to reduce the noise power, we average \eqref{eq:main2simple}  over multiple packet pairs.
Let $\overline{S}_k \eqdef   \frac{1}{M}  \sum_{i=1}^M{ \hat{S}_{i,k}} $ denote the average value of $\hat{S}_{i,k}$ over $M$ packet pairs, and $ \overline{N}_k \eqdef   \frac{1}{M}  \sum_{i=1}^M{ N_{i,k}} $ the resulting average measurement noise. With these positions we 
    finally obtain:
\begin{equation}  
 \Delta_k =  \Delta_k^o  + \mu \cdot c  \cdot  \overline{S}_k  + c  \cdot   \overline{N}_k.
\label{eq:range}
\end{equation}

\subsection{Position estimation}
The output of the previous phase is a set of $K-1$ estimates of differential ranges $\{\Delta_k,  \; k=2 \dots K \}$  between the blind  node and each anchor pair $\left( 1,k \right)$ affected by (unknown) errors $ \overline{N}_k$. 
The following positioning phase (ref. Fig. \eqref{fig:scheme1}) takes in input the set of (estimated) differential ranges and the (known) anchor positions in order to finally estimate of the blind node position.

The position estimation problem can be casted into a Least Square (LS) optimization problem\footnote{Note that he error terms $\overline{N}_k$ are not independent, i.e., their covariance matrix is not diagonal. However, for the sake of simplicity, we neglect here correlations between the error terms  and consider an ordinary (non-weighted) Least Square estimation approach.}. Recall that the LS solution coincides with the  Maximum Likelihood (ML) solution if the measurement noise is normally distributed --- from Fig. \ref{fig:Sijk} this assumption appears to be reasonable  for our dataset.

 The goal is to determine the blind node position $\bm{p}$  that minimizes the 2-norm of the difference between the $K-1$ observations (one for each quadruplet) and the model \eqref{eq:range}, i.e., to compute:
\begin{equation}
\argmin_{\bm{p}}{\sum_{k=2}^K{ \left(\Delta_k   - \Delta_k^o - \mu \cdot c \cdot \overline{S}_k \right)^2 }}
 = \argmin_{\bm{p}}{ \sum_{k=2}^K{ \left( \|  \bm{p} - \bm{p}_k   \| -  \| \bm{p} - \bm{p}_1   \|  - \Delta_k^o - \mu \cdot c \cdot \overline{S}_k \right)^2 }}.
\label{eq:hyp}
\end{equation} 
The actual value of $\mu \eqdef (1+\beta_1)^{-1}$ is unknown, but we can safely neglect it, since the relative error caused by the approximation $\mu \approx 1$   on the final position estimate is negligible, comparable with the value of $| \beta_1| $ --- typically in the order of $10^{-5}$ 
(see proof in Appendix). With this simplification the estimation problem rewrites as: 
\begin{equation}
\argmin_{\bm{p}}{ \sum_{k=2}^K{ \left( \|  \bm{p} - \bm{p}_k   \| -  \| \bm{p} - \bm{p}_1   \|  - \Delta_k^o -   c \cdot \overline{S}_k \right)^2 }}.
\label{eq:hypnew}
\end{equation} 

This is a non-linear minimisation problem to be solved numerically. 
We consider two different resolution methods, hereafter denoted as ``HYP" and ``ILS". 
Both methods follow the same high-level approach based on the iterative resolution of (a sequence of) problem instances representing linearised version of  \eqref{eq:hypnew}. 
However, they differ in the details of how linearisation is achieved.  \\

{\bf Hyperbolic equations (HYP) on differential ranges}.
 This is the method described in
\cite{chan94}, and specifically equations (1)-(7) therein. The problem \eqref{eq:hypnew} is linearized directly via Taylor-series approximation.
The name of the method is due to the fact that each term of the form 
$  \|  \bm{p} - \bm{p}_k   \| -  \| \bm{p} - \bm{p}_1   \| = \text{\em constant} $
 describes an hyperbola with foci in $\bm{p}_k$ and $\bm{p}_1$. \\

{\bf Iterative Least Squares (ILS) on pseudo-ranges}.
This is the standard positioning algorithm adopted in GPS  (see e.g. \cite{ils})  based on an Iterative Least Squares (ILS) procedure run on a set of {\em pseudo-ranges} between the blind node and a set of reference points in known positions (satellites in GPS). The term ``pseudo-ranges" denotes a set of ranges, i.e., distances, determined up to an additive common bias term, formally $\{d_k^* + \delta, \; k=1, \dots, K \}$ where $\delta$ plays the role of an additional variable in the estimation problem.  The ILS algorithm, which was successfully used in a similar context also in  \cite{coluccia14,EUC14}, is able to jointly determine the blind position and resolve for $\delta$. 
In order to apply ILS, we need first to transform the range of differential ranges  $\{\Delta_k, \; k=2, \dots, K \}$  into a set of pseudo-ranges. 
This is achieved by a simple variable substitution $d_1^*  = \delta $, i.e., letting  the distance between the blind node and the reference anchor be treated as an additional independent  variable $\delta$.
In this way, the set of remaining distance variables take the form of a set of pseudo-ranges, i.e.  $\{ d_k^* = \Delta_k + \delta, \;  k=2, \dots, K\}$. 
With the variable substitution  $d_1^* \eqdef \|  \bm{p} - \bm{p}_1  \| = \delta $ the problem  \eqref{eq:hypnew} is first transformed into the following alternative formulation:
\begin{equation}
\argmin_{\delta, \bm{p}}{\sum_{k=2}^K{ \left( \|  \bm{p} - \bm{p}_k   \| - \delta  - \Delta_k^o - c \cdot \overline{S}_k \right)^2 }}
\label{eq:ilseq}
\end{equation}
 then linearised and solved iteratively.  \\

Both HYP and ILS algorithms start from a rough initial guess and improve iteratively at each step by determining the local linear
least-sum-squared-error correction. In our implementation the starting point is set to the center point between all anchors. \\

\begin{figure}
\centering
\subfigure[GARDEN]{\includegraphics[width=8cm]{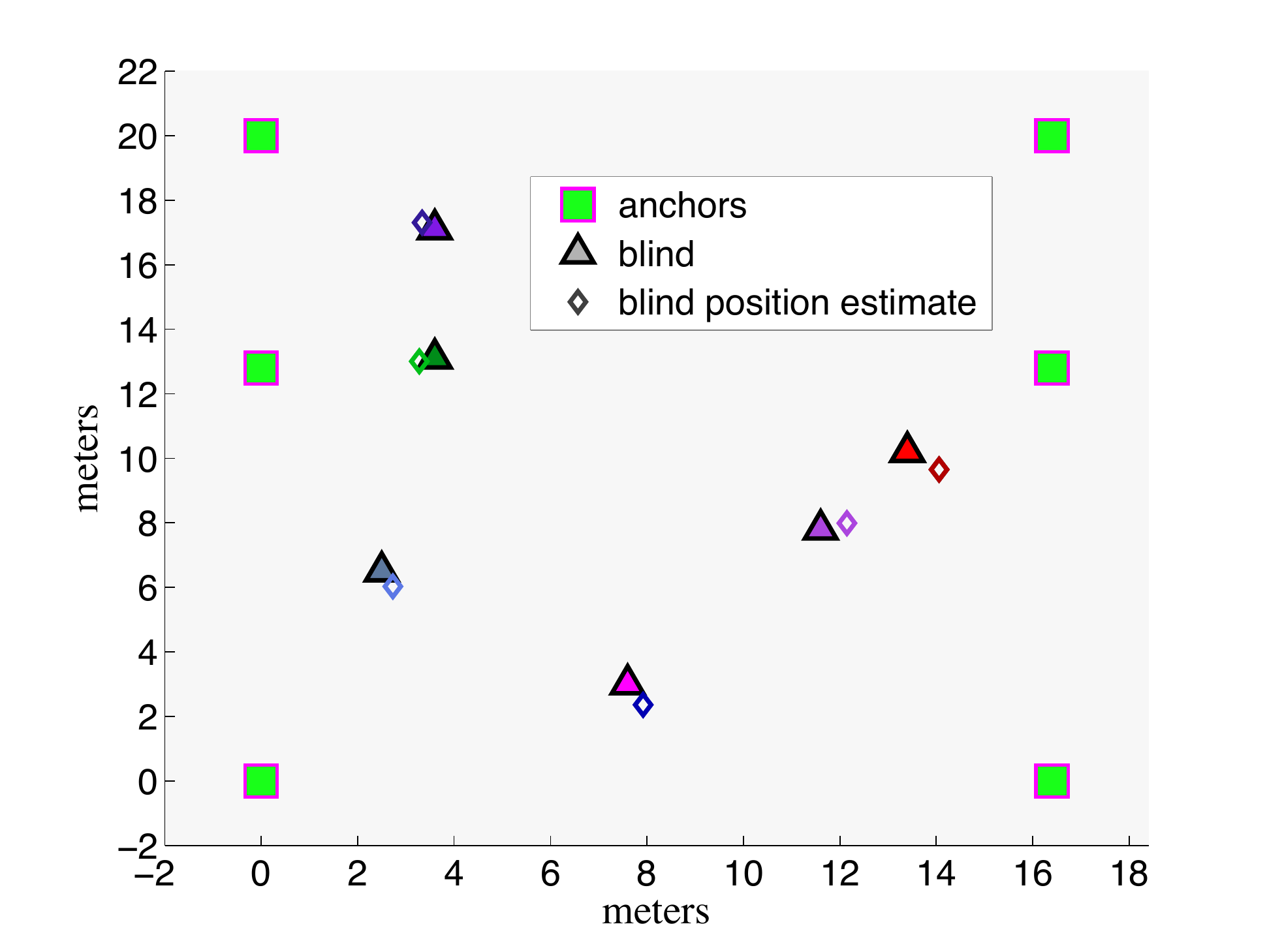}
\label{fig:topo1}} 
\subfigure[GYM]{\includegraphics[width=8cm]{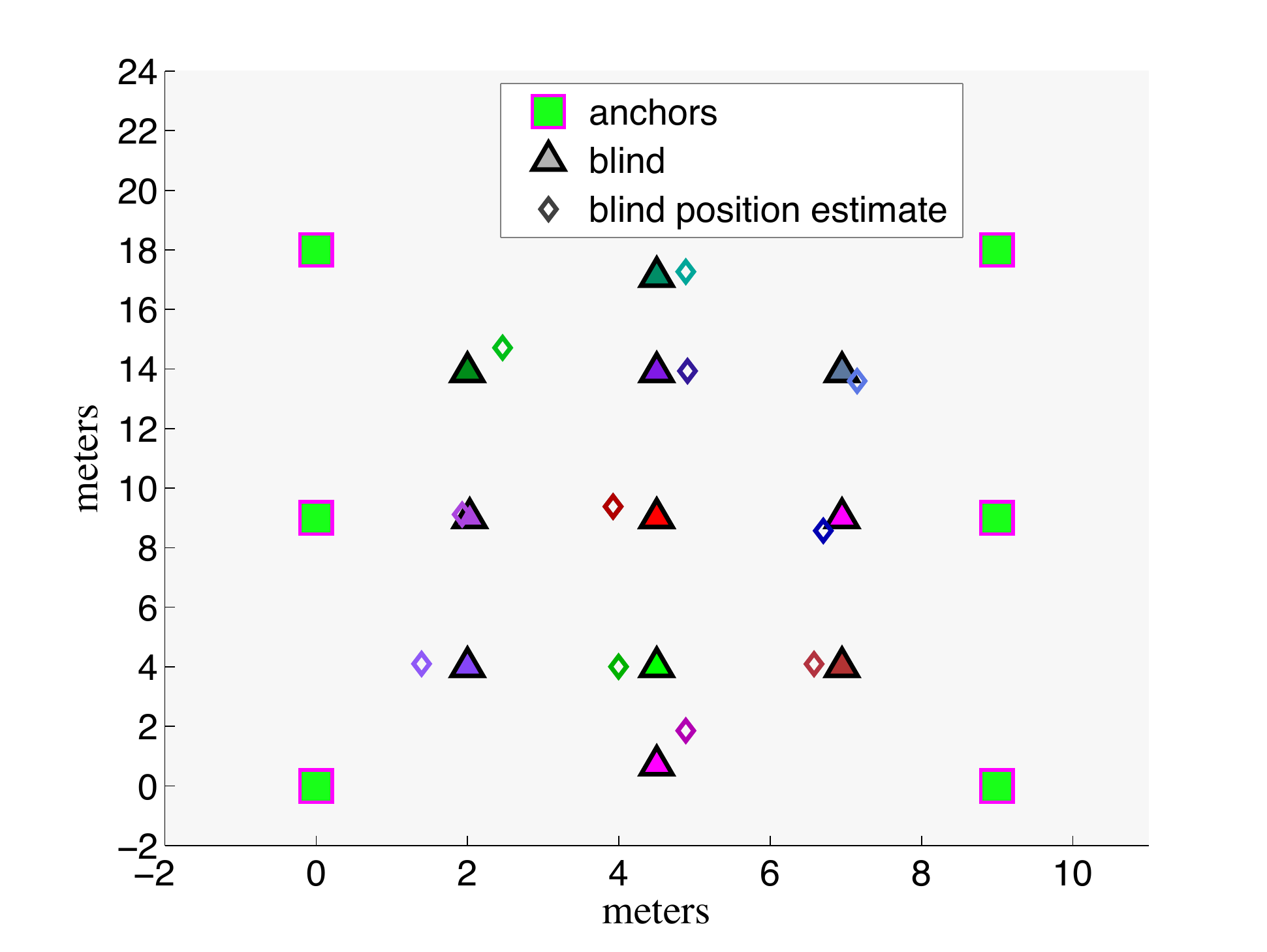}
\label{fig:topo2}}
\caption{Node locations during the experiments. In each scenario, six nodes were kept fixed at locations marked by squares. In each experiment, one node was moved to a different location among those indicated by triangles. The legend refers to the role of each node in the ``active anchors" setting described in Section \ref{sec:active}. }
\label{fig:topo}
\end{figure}

\section{Experimental Results}\label{sec:experiments}
The main challenge is to capture into the model, and properly correct in the algorithm, the non-idealities of the real
timestamping process in standard COTS receivers. 
For this reason it makes sense to validate directly our approach experimentally, with real measurements obtained from a testbed with  COTS devices, instead of resorting to abstract simulations. 


\subsection{Testbed setup}
For our measurement testbed we use WRT54GL devices  from
Linksys. 
This choice was motivated by several reasons: i) their price is very affordable, less then 40
USD on ebay; ii) they run a very robust OpenWRT distribution
based on Linux Kernel 2.6.32; iii) their wireless Network Interface
Card (NIC) is compatible with OpenFWWF \cite{openfwwf}, an open source
firmware that replaces the original binary-only software from
Broadcom and has been widely used as research platform
\cite{wisec14, tmc14, aa2014}. The basic version of OpenFWWF
implements a fully working Distributed Coordination Function (DCF) 
and timestamps each incoming packet with the value of the Basic
Service Set clock,  sampled by the underlying hardware when a
 preamble is successfully received. Unfortunately this clock has only 1 MHz
resolution. For this reason we modified the real-time receiving part
of the code to spin during the reception of specific packets until
they are completely decoded: right after the single instruction
spin-lock,  a sample of the 32 bit  internal clock that
drives the execution pipeline is taken and reported to the Linux
kernel. As a result, we obtain  timestamps with clock frequency of 22 MHz, corresponding to a clock tick of  45.4 ns (13.6 meters at speed light). 
These operations are executed only for UDP packets with
configurable destination port which we check as soon as the first
bytes of each frame are received.

Every node is programmed to transmit UDP packets with a payload of 10 bytes.
Every packet is transmitted with a
single attempt at the MAC layer in order to avoid ambiguities in correlating
 copies of the same packet received by different nodes. In a real scenario this is
equivalent to consider only the first transmission of a frame (retry bit set to 0).
The nodes schedule UDP packet transmissions with an inter-departure period of 10 ms
at the application level. Then, at the MAC layer, the channel access is performed
using the standard DCF protocol with a Contention Window of 15 slots. 
All nodes are set
to transmit in IEEE 802.11g mode with DSSS modulation,  transmission rate of 2 Mb/s and 
long preamble of 192 bits. 
Every  node transmits at an average rate of about
92 packets/second. Every node logs the local reception timestamp, source
address and packet identifier for each received packet. Depending on the role
assigned to each node in each trial (blind, pivot or anchor) we disregard the
packets transmitted or received by that node in that trial.

We conducted two distinct sets of experiments. 
In the first set, all nodes were in full Line-of-Sight (LOS) conditions and all transmissions took  place on a single frequency channel (specifically, channel \#1). 
We tested two different environments: outdoor (garden) and indoor (gym), as described below.  
In each scenario we have used 7 nodes, thereof 6 (anchors and pivot) in fixed positions  that did not change across the different experiments (squares in Fig. \ref{fig:topo}), and one  (blind node) placed in a different position for each experiment. The actual node positions (ground truth) were measured manually with an accuracy in the order of 2-3 centimetres.
A third set of experiments was conducted in mild Non-Line-of-Sight (NLOS) conditions within the main university  entrance hall. The testbed setting and results will be presented  separately in Section \ref{sec:activeNLOS}.

\subsubsection*{First Test Scenario: GARDEN (full LOS)}

The first set of experiments was conducted outdoor in a free area, i.e.,  a green garden. In this test all nodes are placed at $10$ cm height from the ground within a rectangle of  16 $\times$ 20 meters (ref. Fig. \ref{fig:topo1}). All nodes are in Line-of-Sight, without any intermediate obstacle, and the only possible source of multipath error are reflections from the ground.  
Note that in this experiment each node was using two receiving antennas, as per the default configuration. 
We run 6 experiments with different positions of the blind node, indicated by triangles  in Fig. \ref{fig:topo1}. 
 
\subsubsection*{Second Test Scenario: GYM (full LOS)}
The second set of experiments were conducted in a large clear indoor area, i.e., a school gym. 
Nodes were placed at the height of $1.5$ meters from the ground within a rectangle of  9 $\times$ 18 meters (ref. Fig. \ref{fig:topo2}. 
Similarly to the previous test, all nodes are in Line-of-Sight, without intermediate obstacles, and the only possible sources of multipath error are reflections from the floor, from the ceiling and from the perimetric walls, the latter being located at an horizontal distance of 4-5 meters from the nodes. 
Differently from the previous experiments, we removed one of the two antennas from each device in order to enforce single-antenna reception: in fact, if the receiver adopts phase-adjusted combination of the signal from different antennas (e.g. Maximum Ratio Combination), this results in an implicit directionality of the receiving antenna patterns that could in principle increase the risk of reception over Non-Line-of-Sight paths. 
In this scenario we run 11 experiments with different positions of the blind node (triangles  in Fig. \ref{fig:topo2}). 

\begin{figure}
\centering
\subfigure[GARDEN --- HYP]{\includegraphics[width=8cm]{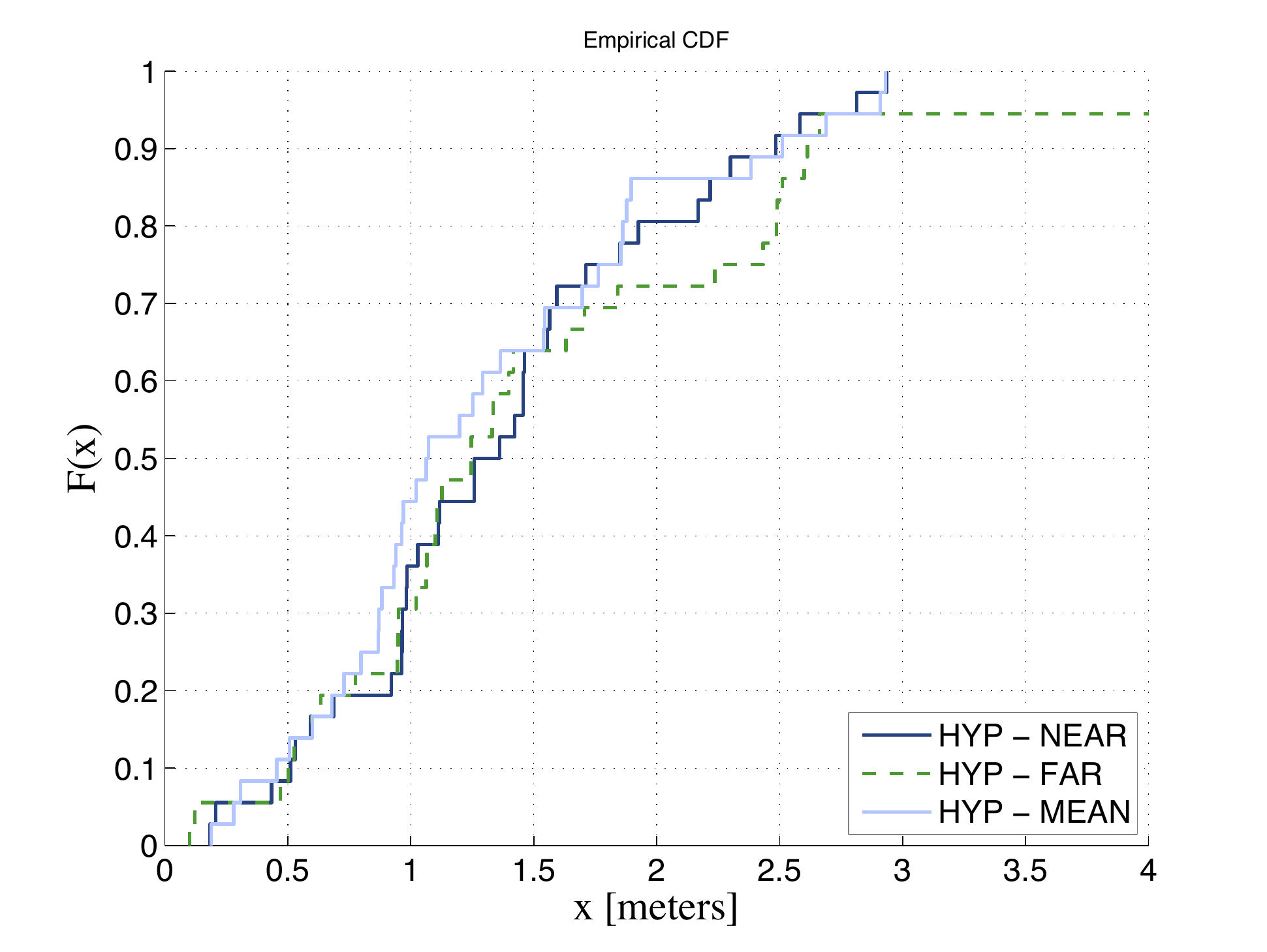}  
\label{fig1_new}} 
\subfigure[GARDEN --- ILS]{\includegraphics[width=8cm]{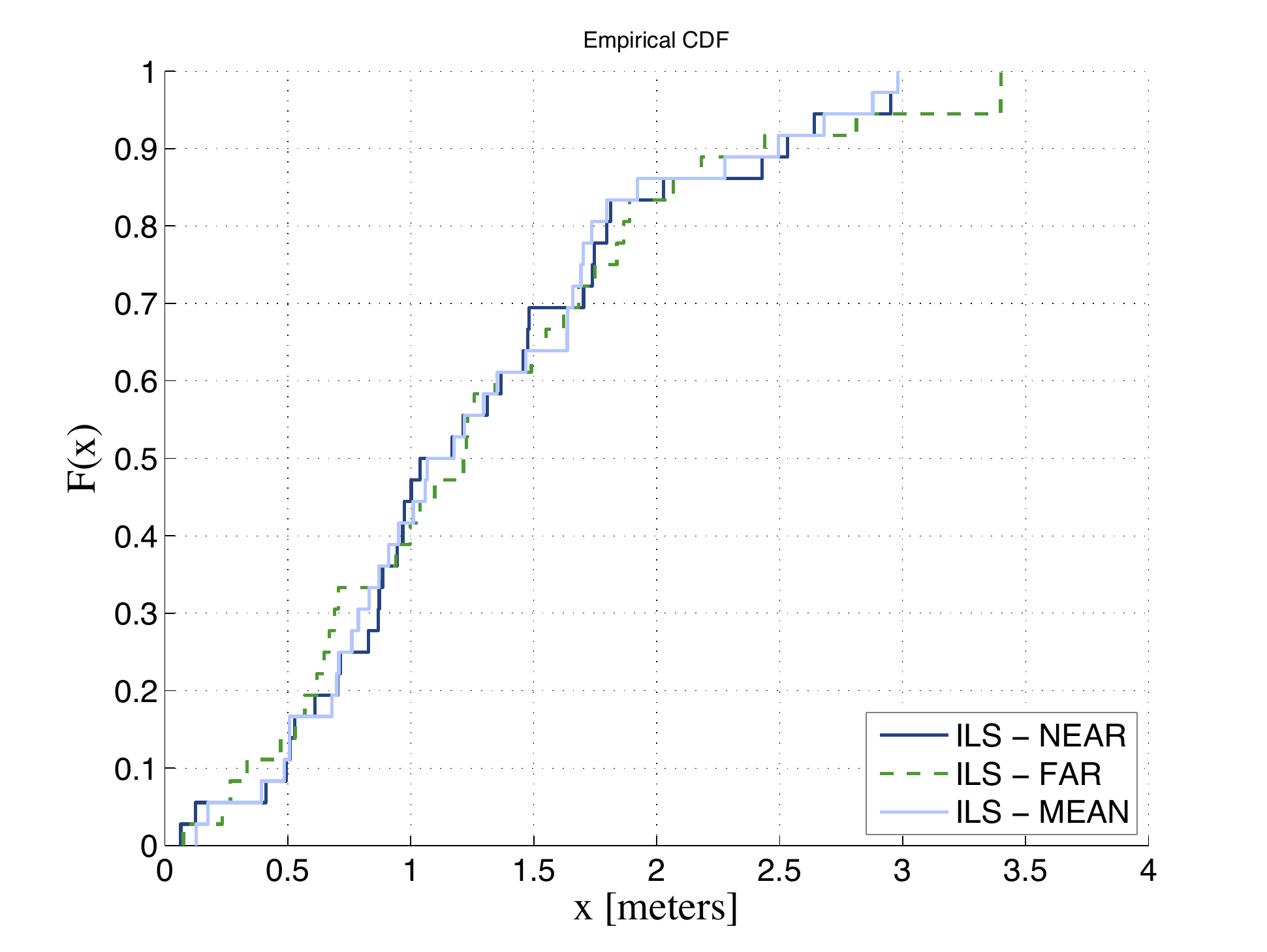} 
\label{fig2_new}}
\subfigure[GYM --- HYP]{\includegraphics[width=8cm]{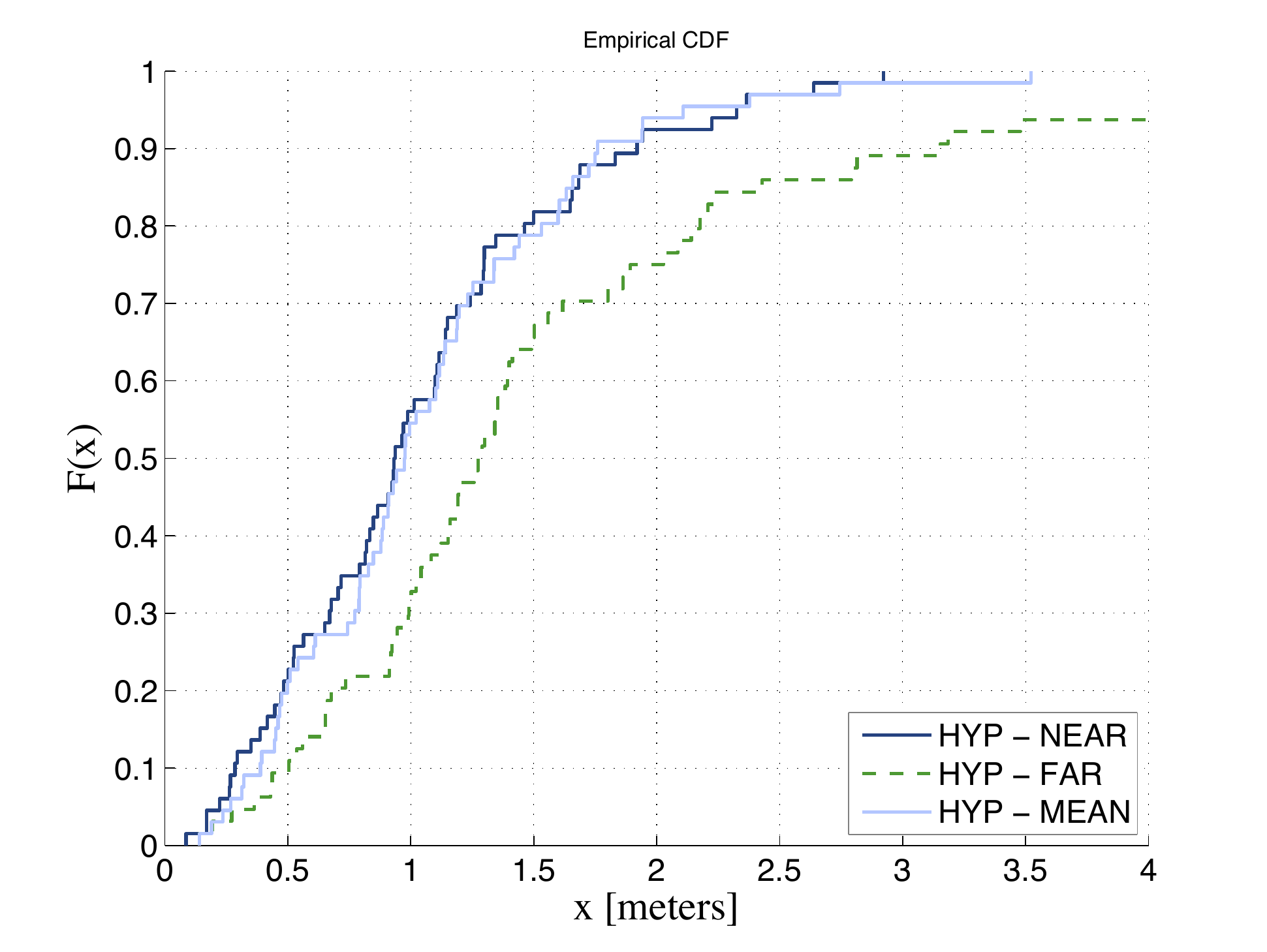}  
\label{fig1_new2}} 
\subfigure[GYM --- ILS]{\includegraphics[width=8cm]{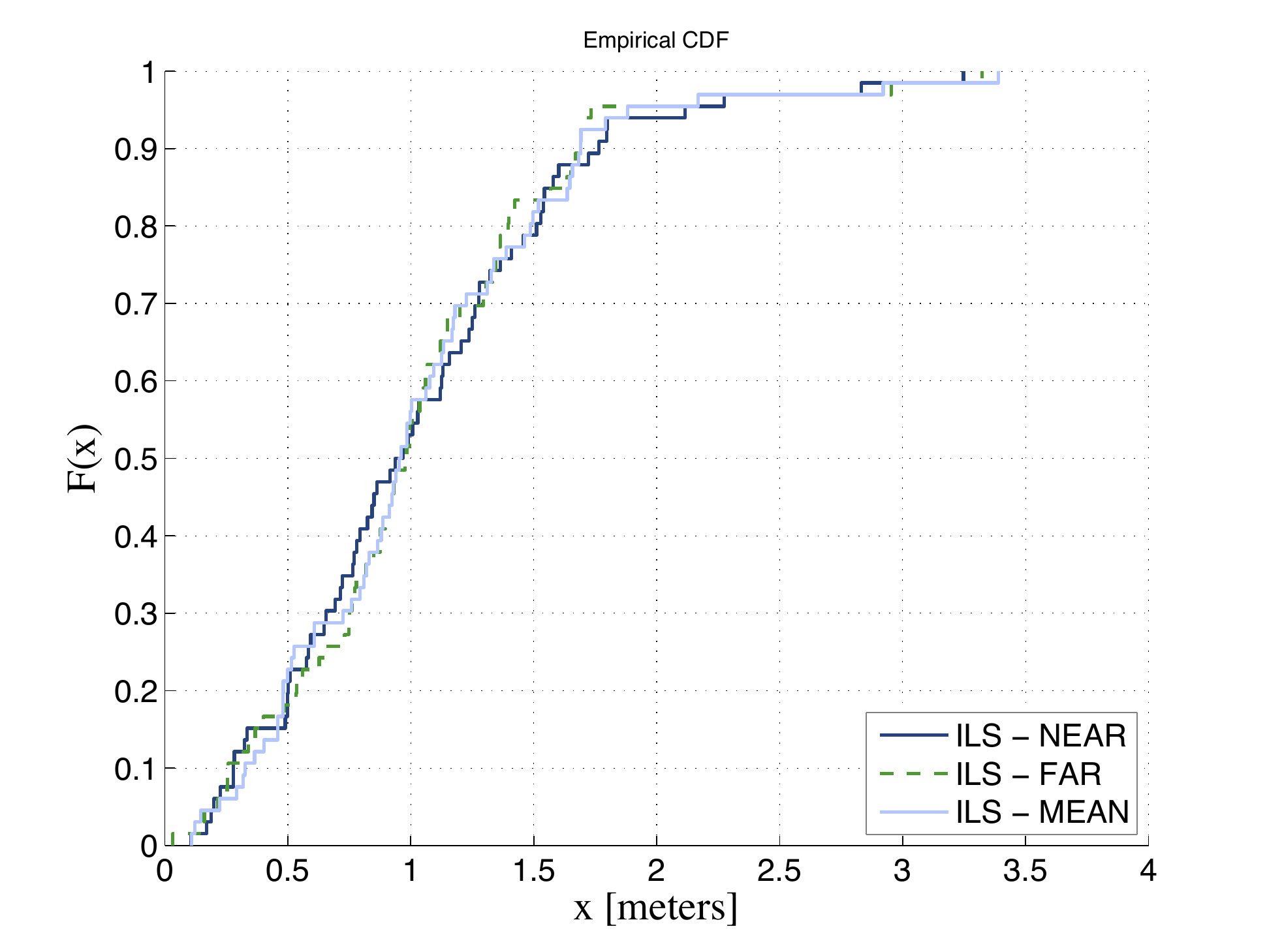} 
\label{fig2_new2}}
\caption{Comparison between HYP (left column)   and ILS positioning (right column), for different choices of reference anchors and $M=200$. }
\label{fig_new}
\end{figure}
 
\subsection{Experimental results with single pivot}

For each experiment, we run 6 different trials by assigning the role of pivot node to each of the fixed nodes, the remaining 5 playing the role of anchors. Therefore, we have a set of 36 different trials (i.e., different configurations of blind, pivot and anchor positions) for the GARDEN scenario and 66  for GYM.  For each trial, we compute the estimated blind position with the algorithm under test, and from there the final position error, i.e., the distance between the estimated and true positions. In order to compare different variants of our method, we plot the Empirical Cumulative Distribution Function (ECDF) of the final position error.

Fig. \ref{fig1_new} and Fig. \ref{fig2_new}  report the ECDF of the position error, respectively, for the two positioning methods HYP and ILS in the GARDEN scenario. 
Similarly, Fig. \ref{fig1_new2} and Fig. \ref{fig2_new2} compare HYP and ILS in the GYM scenario.
The plots are obtained with a single block of $M=200$ packet pairs for each trial, corresponding to a measurement interval of approximately 2 seconds --- recall the packet inter-departure time was set to 10 ms. For each positioning method we have tested three different strategies with respect to the choice of the reference anchor:
\begin{itemize} 
\item {\bf near:}  choose as reference the  nearest anchor to the pivot;
\item {\bf far:} choose the  farthest anchor from the pivot;
\item {\bf mean:} run the algorithm $K$ times,  for all possible choices of reference anchor, and take the mean of the $K$ points.
\end{itemize} 
The plots in Fig. \ref{fig1_new} and  Fig.  \ref{fig1_new2} show that the choice of the reference anchor has a certain impact on the final accuracy when HYP positioning is used, especially in the GYM dataset, and that it is always preferable to  choose the nearest anchor to the pivot. 
The plots  in Fig. \ref{fig2_new} and  Fig. \ref{fig2_new2} show  instead that ILS positioning is pretty insensitive to the choice of reference anchor. 
Overall, ILS and HYP (with near-pivot reference anchor) yield comparable accuracy, with median error around 1-1.2 meter and maximum error around 3 meters. In the remaining of this paper we will use ILS with near-pivot reference anchor, unless differently specified. 

In the next test we are interested to compare two different strategies for combining different subsets of measurements.  
Considering a large dataset of (timestamps associated to) $B$ packet pairs (e.g. $B=200$) collected at $N$ anchors, for a total of $N \cdot B$ timestamps, we consider the following strategies: 
\begin{itemize} 
\item {\bf One-shot.}  The localisation procedure described above is run once, with the whole set of measurements in input. In other words, all  $N \cdot B$  timestamps are processed jointly in a single block.  
\item {\bf Slice \& Prune (S\&P).} With this approach the  set of available measurements  is first ``sliced" into $L>1$ smaller sub-sets, not necessarily disjoint, as explained below. 
The localisation procedure is run independently for each data subset, resulting in a set of $L$ intermediate estimates (data points). An iterative  pruning process is applied onto this set:  take the mean $\overline{\bm{p}}_L$  of all $L$ data points and then prune out the single element point  that is farthest from  
$\overline{\bm{p}}_L$, thus reducing to $L-1$ data points; re-compute the mean $\overline{\bm{p}}_{L-1}$ and iterate the pruning procedure until the number of residual data points reaches a pre-determined stop value $L_{min}$; the mean value $\overline{\bm{p}}_{L_{min}}$ of the residual points is then taken as the final estimate. In our implementation we set $L_{min}=L/2$. This simple pruning procedure is meant to reject data points with occasionally high error, thus reducing automatically to a smaller subset of more consistent data points. 
\end{itemize} 
The initial slicing of input measurements can be performed across one or more of the following dimensions: time, space and frequency:
\begin{itemize} 
\item {\bf Temporal slicing}: a sequence of $B$ packet pairs transmitted on the same channel at different times is sliced into $M$ smaller blocks of $B/M$ packet pairs each.  
\item {\bf Spatial slicing}: from the set of $N$ anchors we can select $ \binom {N} {n} $ subsets  of $n \leq N$  anchors. In order to compute a candidate position estimate, we need at least 3 anchors. Therefore, we consider all possible subsets of 3 or more anchors. The total number of such subsets is $H_{N,3} =  \sum_{n=3}^N{\binom {N} {n}}$. For instance, with $N=6$ we have $H=42$ valid subsets.  
\item {\bf Frequency slicing}: if measurements are performed across multiple frequency channels, 
a natural way of slicing the input data in frequency is to divide them on a per-channel basis. 
\end{itemize} 

\begin{figure}
\centering
\subfigure[GARDEN --- $B=200$]{\includegraphics[width=8cm]{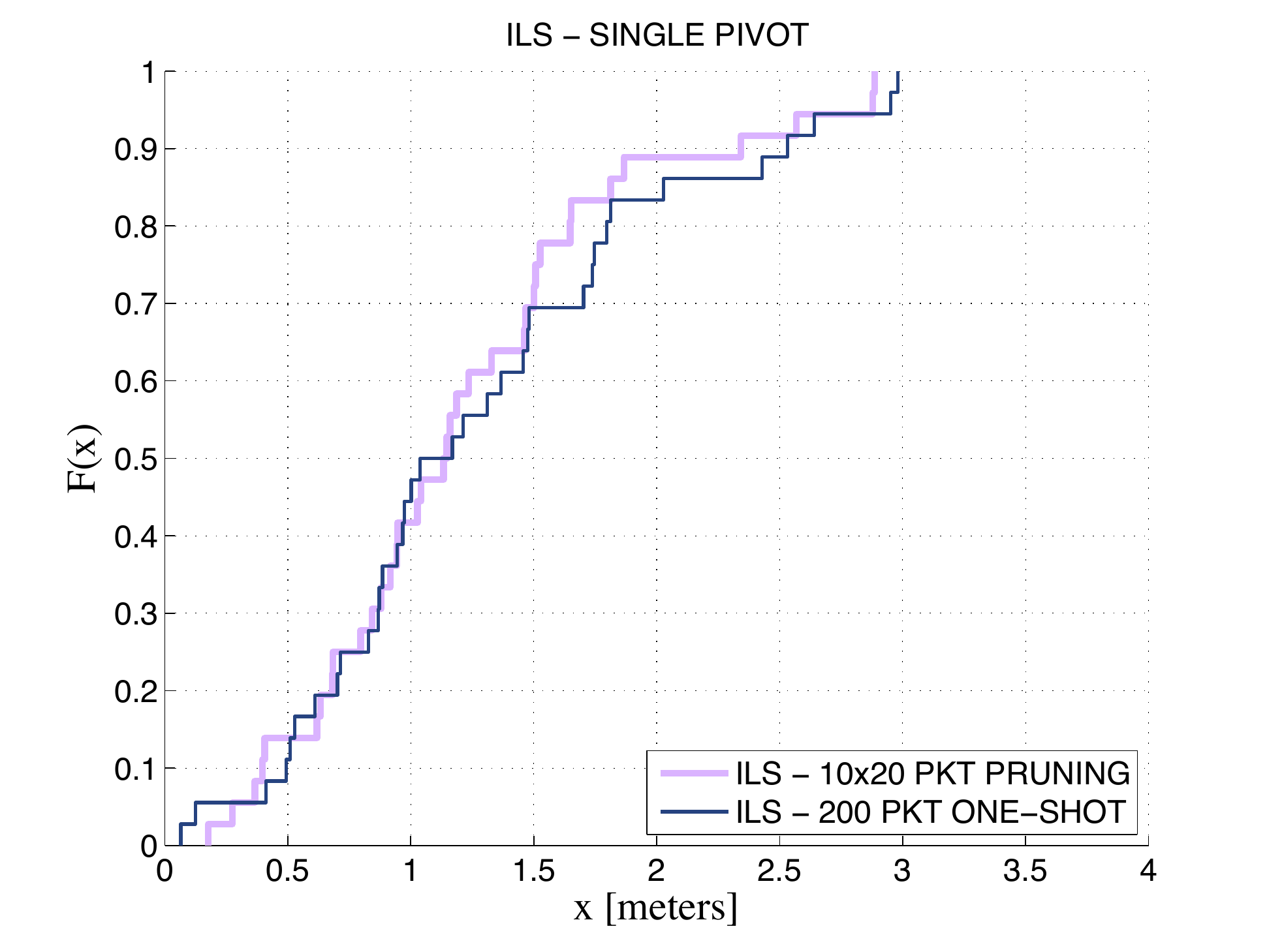}
\label{fig:oneshotvspruning200}}
\subfigure[GARDEN --- $B=1000$]{\includegraphics[width=8cm]{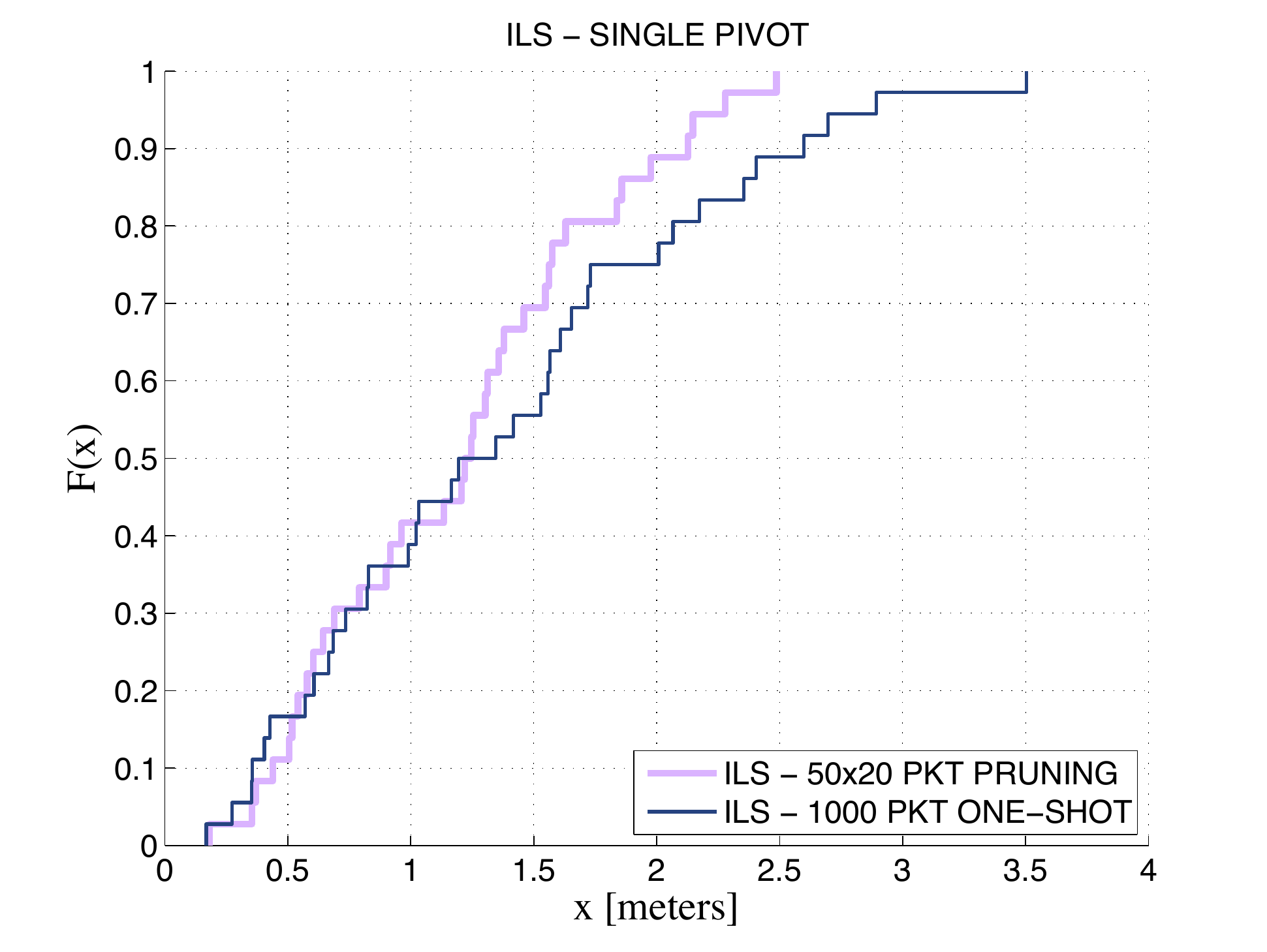}
\label{fig:oneshotvspruning1000}}

\subfigure[GYM --- $B=200$]{\includegraphics[width=8cm]{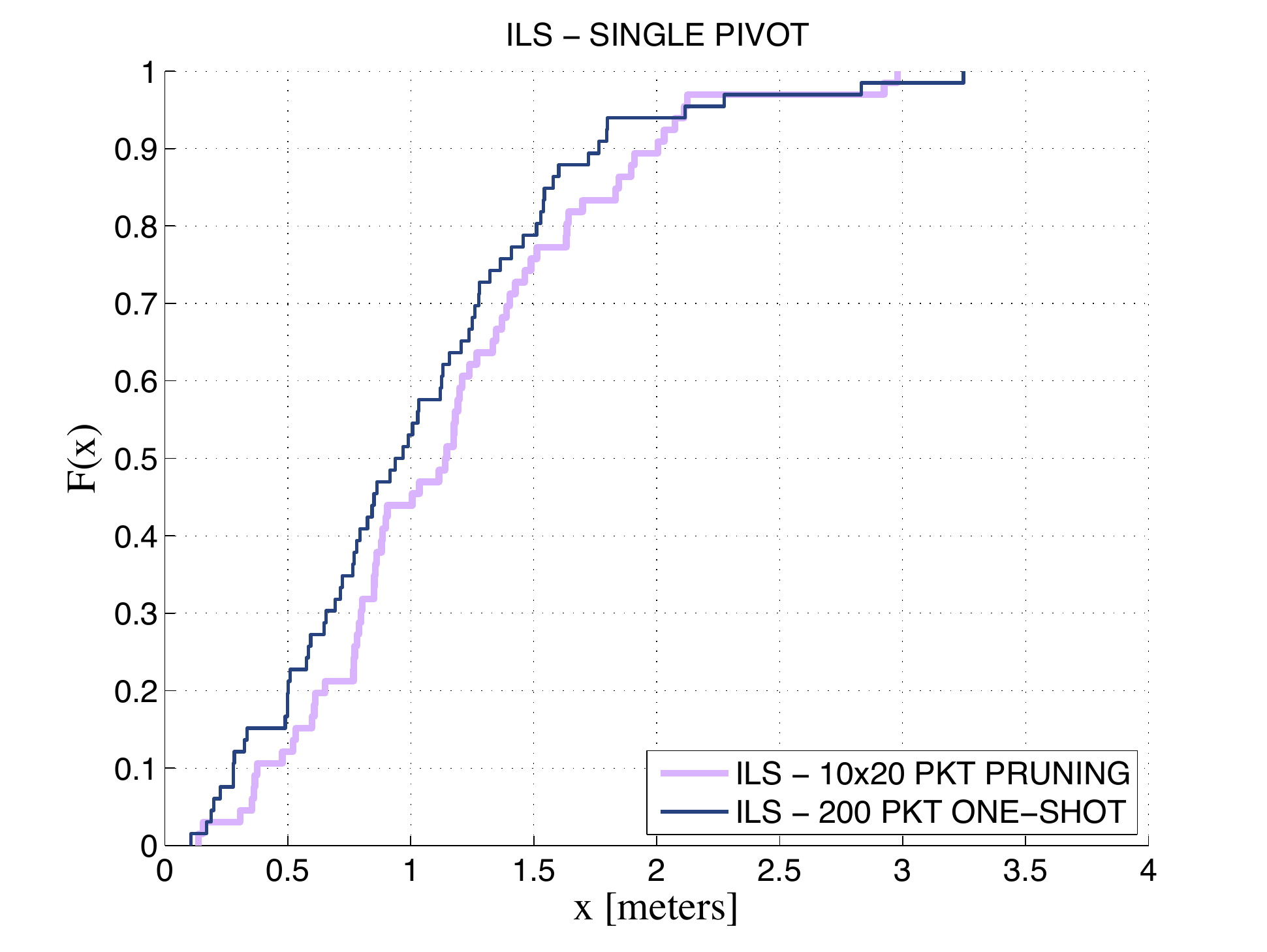}
\label{fig:oneshotvspruning200b}}
\subfigure[GYM --- $B=1000$]{\includegraphics[width=8cm]{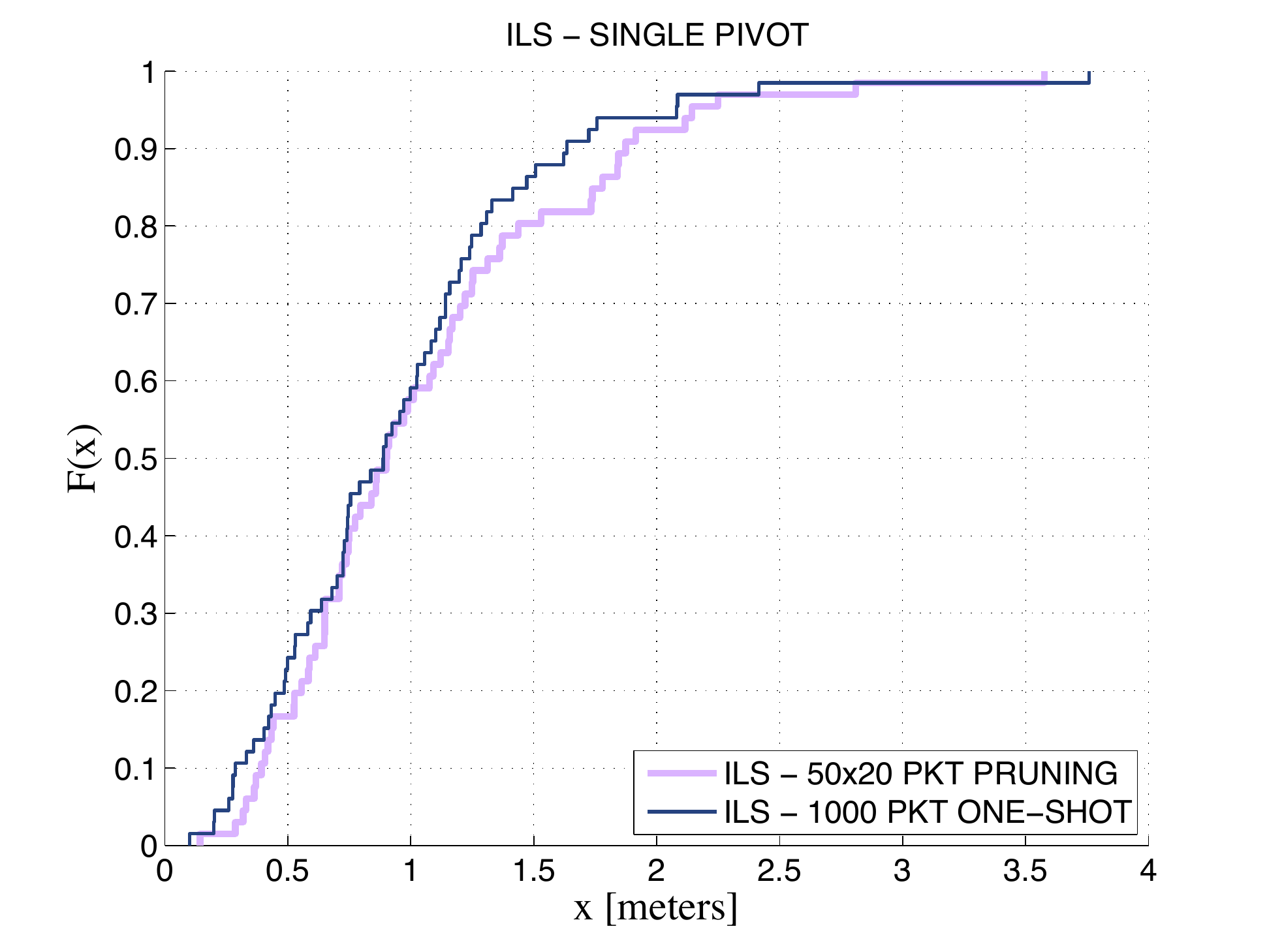}
\label{fig:oneshotvspruning1000b}}

\caption{Comparison ``one-shot"  vs. ``slice \& prune"  for $B=200$ and $B=1000$ (single channel measurements).}
\label{fig:oneshotvspruning}
\end{figure}

In the GARDEN and GYM experiments all measurements were collected on a single channel, therefore frequency slicing does not apply. 
For these two experiments we compared the one-shot strategy against the S\&P with joint temporal/spatial slicing.  
In Fig. \ref{fig:oneshotvspruning} we compare the ECDF of the final error obtained with these two strategies, for $B=200$ and $B=1000$. For the pruning method we have set  $M=20$, $H=42$, $L=H \cdot M$ and  $L_{min} = \frac{1}{2} \cdot L$.  
The first observation  is that the two methods yield comparable results for most of the trials.
In the GARDEN experiments, the S\&P method is able to reduce the error for the few trials with larger error (upper tail of the ECDF). 

 The second observation is that using more packets does not necessarily lead to better accuracy: 
comparing the plots for  $B=200$ and $B=1000$ it can be seen that the accuracy  remains substantially unchanged with S\&P, and slightly degrades with the one-shot method for the GARDEN experiment (note the rightward drift of the upper tail of ECDF). 
The latter finding indicates that the timestamp measurements collected in the testbed are not only affected by a random  component, i.e. measurement noise,  which in general can be reduced by increasing  the number of observations (i.e., packets) but also embed some form of {\em systematic} bias accountable to multipath. 

\section{Extension to active anchors}\label{sec:active}
Until now we have considered a system model with a single transmit-only pivot and multiple receive-only (passive) anchors, as sketched in Fig. \ref{fig:scenario1}.
One obvious  extension is to consider a system with {\em active anchors}, i.e.,  where all infrastructure nodes are capable of both transmitting and receiving (and therefore timestamping) the packets or beacons from other anchors and from the blind node.
This new scenario, sketched in Fig. \ref{fig:scen2},  will become more and more relevant for practical applications as new systems are emerging that involve dense  deployments of many low-cost infrastructure nodes,  e.g.  based on Bluetooth Low-Energy technology \cite{ibeacon} and in WSN applications.

It is straightforward to adapt our algorithm to work in the active anchors scenario.   
The signal reference function, that in the previous scenario was performed by the  pivot node,  can be now absorbed by the anchor themselves:
timestamp adjustment will rely on anchor-to-anchor communications, instead of pivot-to-anchors. 
It is reasonable to expect that the active anchors scenario will allow for higher estimation accuracy when compared with the single-pivot method for the same number and placement of infrastructure nodes, due to the higher {\em volume} of available data (i.e., transmitted packets, hence timestamps) and {\em diversity} of transmitting positions. 

We seek to extend the algorithm presented in the previous section to address the active anchors scenario without adding too much implementation complexity. Motivated by the positive results obtained in the previous experiments
we resort to a procedure based on the same ``slice \& prune" principle described above. 
The key idea is to consider a single scenario  with $N$ active anchors as the overlay of $N$ parallel sub-scenarios, each with 1 transmit-only pivot and $N-1$ receive-only anchors. 
In other words, we are  ``slicing" the problem instance into $N$ parallel instances --- not completely disjoint, as the packets transmitted by the blind node will be common to all sub-instances.
Each of the $N$ sub-instances can be further sliced in space --- considering all subsets of three or more anchors, resulting in $H_{N-1,3} = \sum_{n=3}^N{\binom {N-1} {n}}$ slices  --- and in time --- dividing the block of $B$ packets in $M$ sub-blocks.
By running the basic positioning procedure for each slice of each sub-instance, we produce a set of $L= N\cdot H_{N-1,3} \cdot \frac{B}{M}$ data points, each produced by a different set of input data. The last step is to apply the iterative pruning procedure described above and take the the mean value of the residual $L_{min}$ data points as the final position estimate. 

\begin{figure}
\centering
\subfigure[GARDEN ]{\includegraphics[width=8cm]{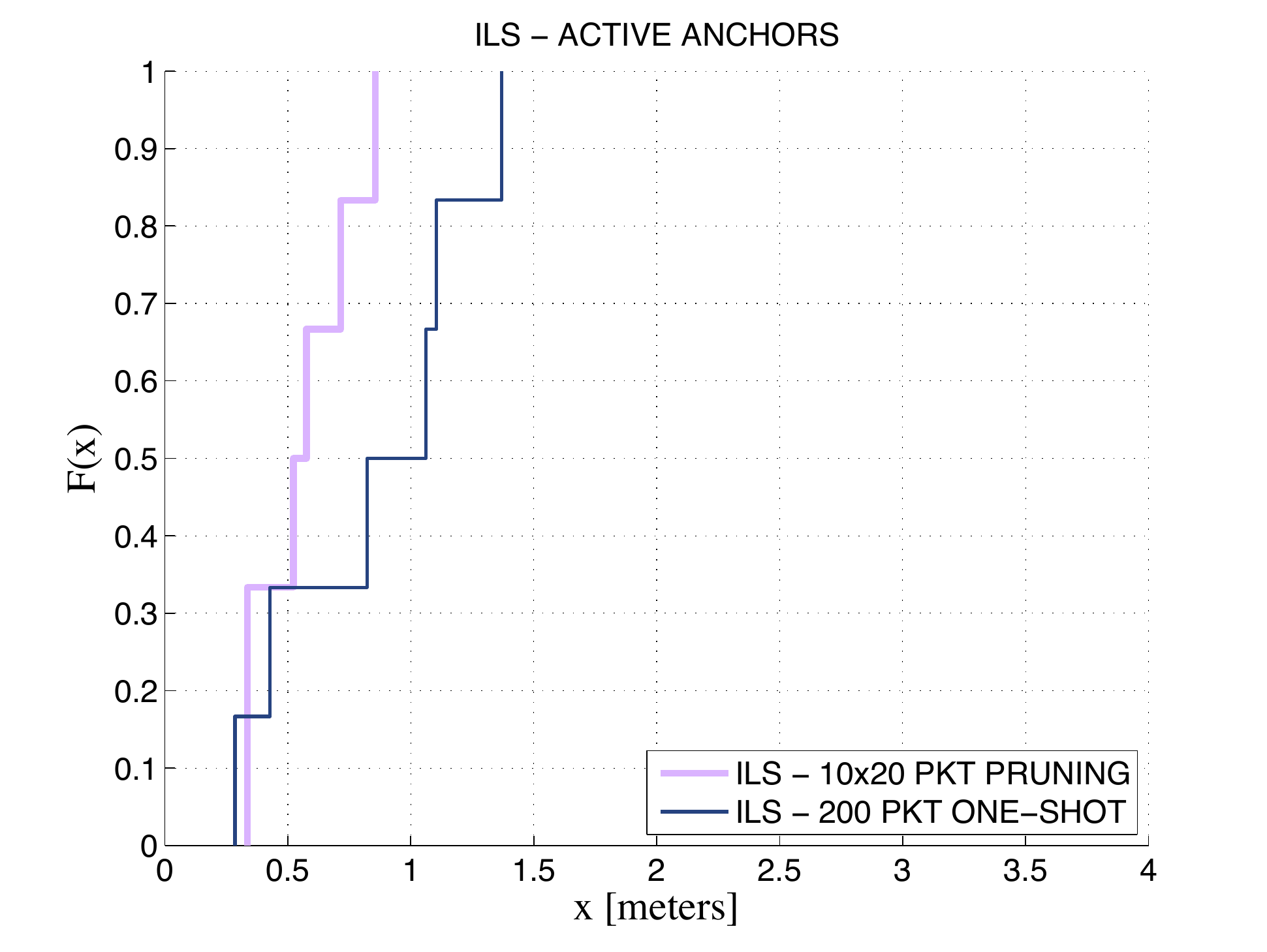}
\label{fig:activegarden}}
\subfigure[GYM ]{\includegraphics[width=8cm]{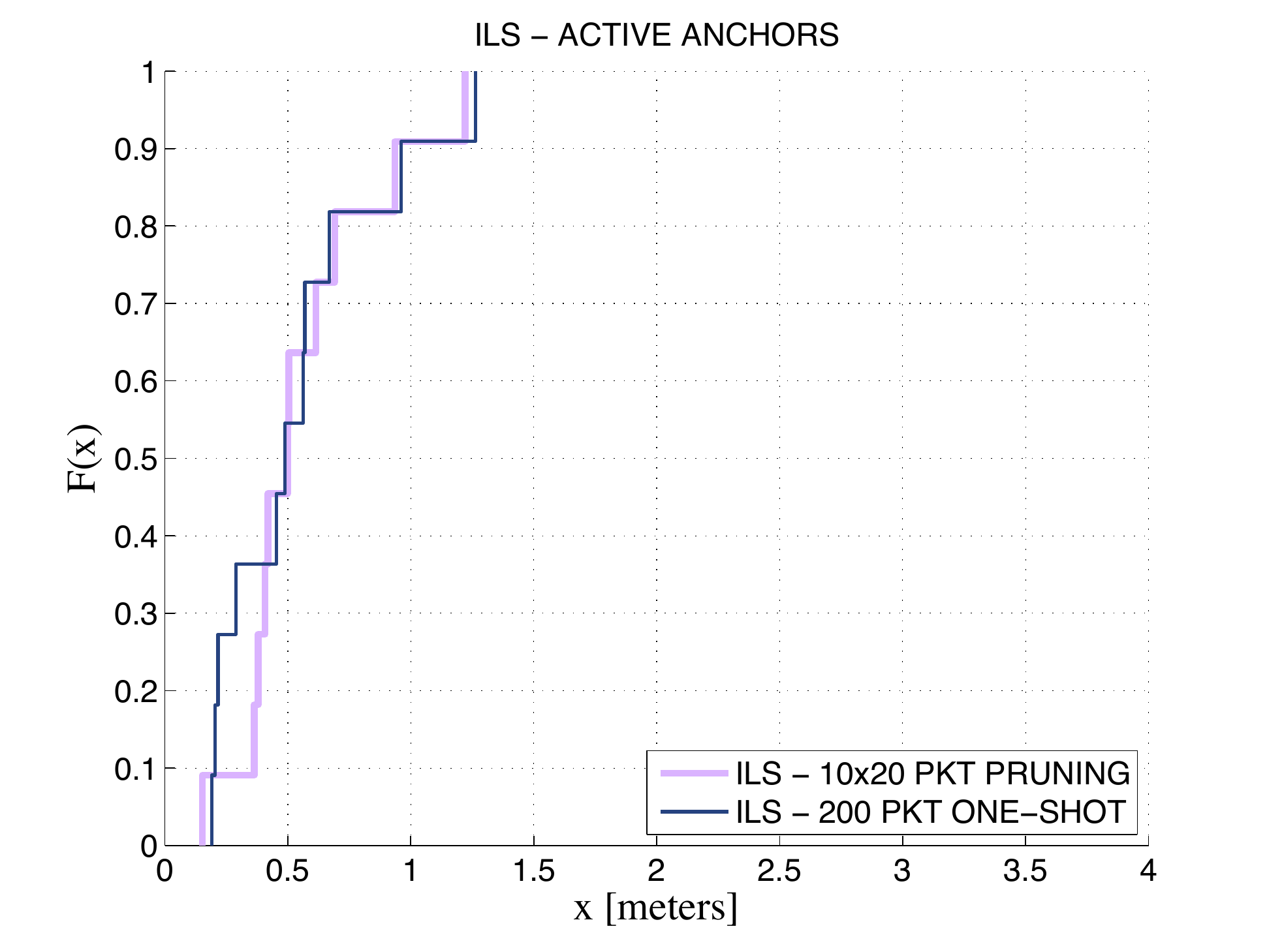}
\label{fig:activegym}}
\caption{Distribution of absolute positioning errors with active anchors. For each experiment the corresponding  blind position estimates are plotted in Fig. \ref{fig:topo}.}
\label{fig:multipivot}
\end{figure}

As done above, we compare this S\&P approach versus a more compact method, where a single datapoint is obtained for each sub-instance by using all $B \cdot (N-1)$ timestamps without slicing,  and the mean value of the $N$ data points is taken directly as final estimate (no iterative pruning).  For each experiment (6 in GARDEN scenario and 11 in GYM) we run the two methods  on the same sample dataset, considering the first $B=200$ packets transmitted   by each node in the given experiment. The final absolute error CDF is reported in Fig. \ref{fig:multipivot}. Consistently with what was observed in the previous test, the S\&P method helps to reduce the final errors in GARDEN (the maximum error reduced from 1.8 to  0.8 meters), while in GYM it has no evident  influence on the accuracy. 
 The final  positions estimated with the S\&P method for   each experiment are also reported as diamond markers in  Fig. \ref{fig:topo}.
 The median error is around 0.5 meter in both environments, and the maximum error is 0.8 meter and 1.3 respectively in GARDEN and GYM.

\section{Robustness to mild NLOS conditions}\label{sec:activeNLOS}
 The results presented above are quite remarkable considering that we have used relatively affordable COTS devices without putting in place any synchronisation  mechanism. However, they were obtained in full LOS conditions, i.e., all nodes (anchors and blind) were in direct visibility to each other. 
In order to  explore the impact of NLOS  on the final accuracy of the proposed method, we conducted an additional series of experiments (``HALL") in a more complex indoor environment, at the main entrance hall of the university building. 

For this new set of experiments we used a total of 10 nodes (9 active anchors plug one blind node) arranged in 6 different topologies.
Seven anchors were deployed in the main entrance hall while the remaining two anchors were positioned in nearby corridors, without direct visibility to all other nodes. Note that all node pairs could successfully exchange packets, i.e., NLOS links were not in outage. 
This resulted in a mixture of LOS and NLOS links,  with prevalence of LOS links. In other words, we reproduced ``mild NLOS" conditions according to the tassonomy proposed in \cite{vaghefi13nlos}. To counteract the NLOS error we leveraged \emph{frequency diversity}: different blocks of packets were transmitted on four different radio channels --- specifically channels \#1, \#5, \#10 and \#14 of IEEE 802.11g ---  and frequency slicing was adopted along with temporal and spatial slicing in S\&P. 

\begin{figure}
\centering
\includegraphics[width=9cm]{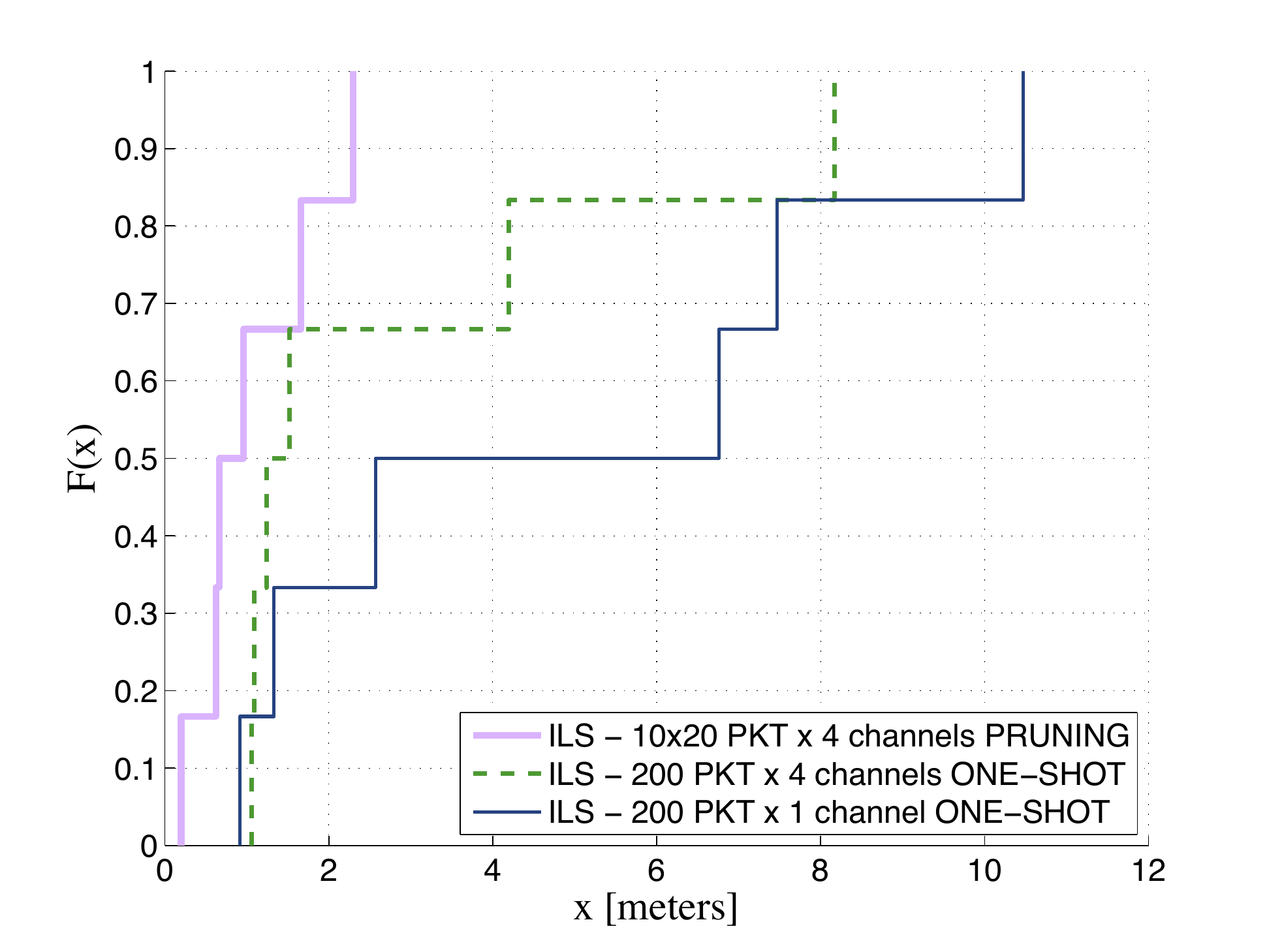}
\caption{Distribution of absolute positioning errors for HALL experiments (6 trials) in mild NLOS conditions.}
\label{fig:nlosresult}
\end{figure}

The results are reported in Fig. \ref{fig:nlosresult}. The rightmost curve shows the final error obtained by the one-shot estimation with data from a single channel. 
This curve should be compared with the one-shot curves in Fig. \ref{fig:multipivot} that were obtained in similar setting (block of $B=200$ packet pairs transmitted on single-channel and processed in one-shot), the only difference being (i) the number of anchors and (ii) the presence of NLOS link. Despite the HALL experiments were performed with more anchors than GYM and GARDEN (9 against 7), the one-shot approach can not prevent the final positioning error to increase dramatically, due to NLOS error in input: the worst case error rises from less than 1.5 meters in GYM/GARDEN (ref. Fig. \ref{fig:multipivot}) to more than 10 meters in HALL. 
The middle curve in Fig.  \ref{fig:nlosresult} was obtained with multi-channel measurements over 4 different channels, for a total of $4 \cdot B= 800$ packet pairs, processed together in a single round. Compared with the previous curve, it can be seen that the introduction of frequency diversity (multi-channel measurements)  has a limited positive impact on the final error  --- the maximum value reduces to 8 meters.  

The leftmost curve was obtained by running the S\&P algorithm on multi-channel data. Slicing was applied over all three dimensions --- time, space and frequency. The coupling of S\&P and frequency diversity leads to a dramatic improvement, with a reduction of the maximum error down to 2 meters, and a median error of only 1 meter, not much higher than what was obtained in GYM and GARDEN in full LOS conditions. 
These three curves demonstrate two things. First, that the basic  DTDOA with one-shot estimation is extremely sensitive to NLOS, and even a small amount of NLOS links has devastating effect on the final accuracy. Second, that combination of S\&P and multi-channel measurements is successful in rejecting (a limited amount of) input NLOS errors, enabling the practical adoption of DTDOA in mixed LOS/NLOS environments.  

\section{Discussion and outlook on future work} \label{sec:discussion}

\subsection{Improving robustness in heavy NLOS scenario}
In practical settings the most serious source of error is NLOS propagation. 
Following \cite{vaghefi13nlos} we  discriminate between ``mild NLOS" and "heavy NLOS" environments, depending on whether the LOS links are prevalent or not.  
Our experimental results were obtained in full LOS (GYM and GARDEN experiments) or mild NLOS conditions (HALL experiments). 
In the latter case, we managed to maintain satisfactory results by leveraging frequency diversity --- in addition to spatial and temporal diversity --- via the S\&P method with multi-channel measurements. We remark that the S\&P method plays  a key role in rejecting a small amount of NLOS error (recall from Fig. \ref{fig:nlosresult} that simple averaging over multiple channels does not achieve the same improvement). However, our method will simply not work satisfactorily in heavy NLOS environments.

Generally speaking, the NLOS problem affects all radio  localisation methods, whether RSSI-based or time-based, including the traditional ToA, TDoA and of course also the  DTDOA method considered here. In other words, it should not be considered as problem specific to our method. However, NLOS errors propagate in different ways within  ToA, TDoA and DTDoA localisation procedures. Therefore, the great deal of proposed solutions for NLOS mitigation that have been developed for ToA and TDoA 
(see \cite{guv09,vaghefi13nlos,hara13} and references therein) will probably need careful adaptation before  they can be applied to DTDoA systems.
For this reason, further work is needed to understand the sensitivity of DTDoA to NLOS errors and to develop NLOS mitigation strategies specifically for DTDOA. One important direction in this regard, for DTDOA as well as for other time-based techniques,  is to leverage physical layer information, from simple RSSI up to Channel State Information (CSI) and  Angle-of-Arrival (AoA) as done e.g. in \cite{Sen13} with IEEE 802.11n nodes.  However,  this approach comes at the cost of reduced generality, as it imposes additional requirements on the transceiver devices (e.g. multiple antennas with synchronous receiver chains,  and firmware support for CSI exporting) and implicitly rely on specific radio standard (e.g. IEEE 802.11n). In other words, exploiting PHY information involves a trade-off in the applicability space:   it widens the range of environments where time-based localisation can be satisfactorily adopted (from pure LOS and mild NLOS environments towards heavy NLOS spaces), at the cost of shrinking the range of transceiver models and radio standards  that can be used for this purpose.

\subsection{Distributed and secure computation}
In this work we have considered a centralised approach, where a single entity (e.g., a localisation server or cloud service) concentrates all the data and computation. However, owning to the fact that the basic algorithm involves simple linear operations,  it should be possible to devise methods to perform the same computation in a distributed fashion, without the need of data centralisation. Another related research direction would be to explore the possibility of performing  the localisation procedure without revealing some of the input data,  that might be considered sensitive by the participating entities, e.g., the (known but private) anchor node positions, by drawing methods from the field of  privacy-preserving  computation and  Secure Multiparty Computation.

\section{Related work}\label{sec:related}
The vast majority of previous work on time-based localisation in asynchronous networks (e.g. \cite{wang11,vaghefi13icassp,kim2014,Liu10,Zach14,Gholami13,Li06})  rely on the knowledge or tight control of transmission times (in addition to reception times) and anyway require some form of involvement by the blind node. For these reasons,  such methods can not be used to implement  fully opportunistic localisation on top of legacy wireless systems with COTS hardware, and therefore can not be directly compared to our work. 

The  method elaborated in this paper relies on the (double) subtraction of four arrival times referred to a quadruplet of nodes --- two transmitting and two receiving, ref. equation \eqref{eq:finaldiff}   --- in order to get rid of the unknown transmission times and clock offsets. 
This basic principle characterises an entire class of techniques that are commonly referred to as Differential Time-Difference of Arrival (DTDoA)  --- to  distinguish  from the classical  ToA and  TDoA approaches that require synchronisation between all nodes (ToA) or between the anchors (TDoA). 
Considering the relative ease of adoption and  the possibility to ``opportunistically" implement DTDoA in legacy systems, it is a bit surprising that DTDoA techniques have attracted very small attention  by the wireless networking community until now, at least when compared to the much more popular TDoA methods that, conversely, require to put in place additional burdensome synchronisation mechanisms.  
In this section we review previous work on DTDoA methods and their experimental evaluation, highlighting the main differences with our work.

The DTDoA principle is reminiscent of the ``double-difference" method applied in GNSS to eliminate  residual errors on the carrier \emph{phase} for high-precision applications (see e.g. \cite{tiberius08}).  Considering the problem of terrestrial (non GNSS) localisation with time (not phase) measurement, the DTDoA 
principle first appeared a decade ago in the context of interferometry \cite{maroti05} and was later considered in  systems where the  time difference between two terrestrial signals from distinct transmitters is measured by waveform correlators  \cite{winkler06,fan07,xu13,Leng14}.  This approach demands specialised processing at the physical layer and/or considerable additional capacity to exchange sampled waveforms between the nodes, therefore cannot be implemented on top of legacy devices and systems. 
The idea of using simple reception timestamps (as opposite to cross-correlation and interferometric measurements)  in a DTDoA context was considered in \cite{fan08,exel11}. 
The work in \cite{fan08} addresses the problem of localising a set of \emph{receivers} (from legacy TV and AM broadcast signals), while here we focus on the opposite problem of localising a \emph{transmitter} (of WiFi or other packet-based protocol).

The recent work  \cite{exel11}  is certainly the closest one to our work: the system model considered therein  is equivalent to our single-pivot scenario, and the localisation procedure described there is similar to the basic scheme proposed here with passive anchor. However, there are a number of  differences between the two methods. First, we obtain (an estimate of) the relative clock skews (factor $\gamma_k$) by linear regression over a block of multiple packets, instead of subtraction between consecutive packets and filtering as done in  \cite{exel11}. This allows for a more robust and simple estimation. Second, we adopt ILS positioning instead of hyperbolic equations, which (as we have shown) delivers slightly more stable results. Third, we combine the basic method with a  novel  `` slice \& prune"  strategy that, on one hand, adds robustness by rejecting occasional  ``bad" data and, on the other hand,  provides a very simple approach to effectively fuse data from different transmitters (pivot nodes) if available. 
In this way, it is straightforward to apply our localisation procedure to the active anchors scenario, which was not considered in  \cite{exel11} and whose relevance for practical applications is probably going to increase in the near future, following the widespread of wireless systems with dense deployment of  low-cost fixed nodes, e.g. based on Low-Energy Bluetooth  \cite{ibeacon} and in WSN.
Note also that the experimental results  reported by \cite{exel11} are limited to 1D localisation, i.e., the blind node is constrained on a linear axis, while in our experiments we consider 2D localisation. Finally, it should be mentioned that the testbed results reported in \cite{exel11} were obtained with a self-developed transceiver device rather than COTS hardware. 
To the best of our knowledge, the present paper is the only work to report sub-meter  accuracy for 2D localisation from experimental measurements obtained exclusively with unsynchronised  COTS devices.  

For the sake of completeness we cite also the work by Coluccia et al. in \cite{coluccia14}  and \cite{EUC14}  where  the problem of time-based localisation in asynchronous networks is addressed from a slightly different perspective. 
While DTDoA methods eliminates both unknowns, namely the transmission time and the clock bias, by {\em double subtraction} between four link equations for a {\em quadruplet} of nodes  (two transmitters and two receivers), Coluccia et al. resort to {\em single subtraction}  between  two link equations for a {\em triplet} of nodes, namely two transmitters and one receiver \cite{coluccia14},  or one transmitter and two receivers \cite{EUC14}. 
In this way, they eliminate only one of the two unknowns  --- namely the clock bias in \cite{coluccia14} and the transmission time in \cite{EUC14} ---  and  treat the other unknown as a nuisance parameter to be estimated jointly with the (pseudo)-ranges. However they do not consider clock drifts in their model: this is a serious limitation for the applicability of their methods in practical scenarios where clock frequency deviations can not be ignored --- in fact  \cite{coluccia14,EUC14}  provide only simulation results, not experimental validation. We believe however that future extensions of those work in the direction of a more realistic clock error model bear the potential to improve further the achievable localisation and possibly beat our method.

\section{Conclusions} \label{sec:conclusions}
We have proposed and demonstrated a radio-based localisation method that relies exclusively on the reception timestamps collected by multiple anchors for the packets transmitted by the blind node and (at least one) other emitting node(s)  in known position(s). Differently from  traditional time-based localisation schemes,  our method does not require any form of node synchronisation, nor the measurement (or control) of transmission timestamps. Also, it does not require the collaboration of the blind node, that can be localised while performing standard communication functions. The proposed method can be run ``opportunistically" on legacy signals and packets transmitted over-the-air for communication purposes, regardless of the signal format and protocol, and in principle can be applied to any wireless technology (including WiFi, Bluetooth, IEEE 802.15.4, DVB-T, UMTS, LTE, etc.)  and any combination thereof.

In principle, the  system model can be applied to ``cooperative localisation" scenarios where some of the (transmitting and/or receiving) mobile nodes know their position only approximately, e.g., from their onboard GNSS module. These nodes can participate to the localisation procedure by providing timestamp measurements (as receivers) and/or reference signals or packets (as transmitters), similarly to the scenario considered already in \cite{coluccia14}.  The (unknown) error in the initial position of these nodes represents an additional source of measurement error that, however, does not break the conceptual formulation of the problem. The joint processing of these data (initial positions and reception timestamps)  not only provides a position estimate for the blind node(s), but bears also the potential of improving the initial position estimate delivered by GNSS modules. In other words, it can serve as a ``cooperative augmentation" strategy for GNSS positioning that, differently from previous proposals \cite{gar12,ami14},  does not require the  implementation of additional ranging functions and sensors but can exploit opportunistically existing communication signals and devices. 

The experimental results presented in this work indicate that sub-meter accuracy can be reached nowadays with low-cost COTS hardware, at least in LOS conditions. We believe that the combination of minimal implementation requirements and good achievable accuracy make this approach very appealing for practical applications in a wide range of application scenarios,  most prominently those characterised by a high density of radio devices and signals, e.g., for indoor positioning in crowded areas (stations, airports, shopping centres) and in C-ITS  applications. 

In the progress of the work we aim at exploring more advanced estimation methods for the problem at hand that can be naturally extended to cope with mobile nodes (blind and/or anchors), thus shifting the focus from the problem of static positioning towards dynamic tracking.   
Another important direction for further work is to develop robust NLOS mitigation strategies tailored for the scenario at hand.  

\appendix
\section{Proof of insensitivity to a common skew term.}
Hereafter we  proof that fixing   $\mu \eqdef  (1+ \beta_1)^{-1} \approx 1$ in the position estimation problem   \eqref{eq:hyp} leads to an error (in meters) comparable to  $| \beta_1| \ll 1$, whose typical values are in the order of few ppm. In fact \eqref{eq:hyp} can be rewritten as:
\begin{equation}
\begin{split}
&  \argmin_{\bm{p}}{ \sum_{k=2}^K{ \left( \|  \bm{p} - \bm{p}_k   \| -  \| \bm{p} + \bm{p}_1   \|  -  \|  \bm{p}^o - \bm{p}_k   \| +  \| \bm{p}^o - \bm{p}_1   \|   -  \mu \cdot c \cdot\overline{S}_k \right)^2 }}  \\
& \quad =  \argmin_{\bm{p}}{  \frac{1}{{\mu^2}}  \sum_{k=2}^K{ \left( \|  \bm{p} - \bm{p}_k   \| -  \| \bm{p} - \bm{p}_1   \|  -  \|  \bm{p}^o - \bm{p}_k   \| +  \| \bm{p}^o - \bm{p}_1   \|   -  \mu \cdot c \cdot \overline{S}_k \right)^2 }}  \\
& \argmin_{\bm{p}}{    \sum_{k=2}^K{ \left( \frac{1}{\mu}  \|  \bm{p} - \bm{p}_k   \| - \frac{1}{\mu}  \| \bm{p} - \bm{p}_1   \|  -  \frac{1}{\mu}  \|  \bm{p}^o - \bm{p}_k   \| +  \frac{1}{\mu}  \| \bm{p}^o - \bm{p}_1   \|   - c \cdot  \overline{S}_k \right)^2 }}.
\end{split}
\label{eq:app}
\end{equation} 
For a generic position vector $\bm{p} = \left[ x,y\right] $ in the standard reference system denote by $\bm{q} \eqdef  \frac{1}{\mu} \cdot \bm{p} = \left[ \frac{x}{\mu},\frac{y}{\mu}\right] $ the corresponding vector in a new reference system rescaled by $\frac{1}{\mu}$. Note that the distance between two point vectors in the new system is also rescaled by the same factor, i.e., $\| \bm{q}_1 - \bm{q}_2  \| = \sqrt{\left( \frac{x_1-x_2}{\mu} \right)^2 + \left( \frac{y_1-y_2}{\mu} \right)^2} = \frac{1}{\mu} \| \bm{p}_1 - \bm{p}_2  \|  $, therefore the problem \eqref{eq:app}  is equivalent to:
\begin{equation}
 \argmin_{\bm{q}}{ \sum_{k=2}^K{ \left( \|  \bm{q} - \bm{q}_k   \| -  \| \bm{q} - \bm{q}_1   \|  -  \|  \bm{q}^o - \bm{q}_k   \| +  \| \bm{q}^o - \bm{q}_1   \|   - c \cdot  \overline{S}_k \right)^2 }}, 
\label{eq:app2}
\end{equation} 
where all vector points are expressed in the new rescaled system. 
The solution $\bm{q}^*$ of \eqref{eq:app2} corresponds to  the  solution $\bm{p}^*$ of  \eqref{eq:app} expressed in the new reference system. 
Since the actual value of $\mu$ is unknown, we cannot  exactly map the point $\bm{q}^*$ back to  the original reference system, i.e., we cannot derive $\bm{p}^*$ exactly.  However, the relative error $e_{\mu}$ caused by ``interpreting" the coordinates of any generic point  $\bm{q}$ in the original reference system is equal to $|\beta_1|$:
\begin{equation}
e_{\mu,\bm{p}} \eqdef \frac{  \| \bm{q} - \bm{p}  \| }{ \|   \bm{p}  \|  }=  \frac{  \| \frac{1}{\mu} \cdot \bm{p}   - \bm{p}  \| }{ \|   \bm{p}  \|  } =  \left| \frac{1}{\mu} - 1 \right|  = |\beta_1|.
\end{equation}
Q.E.D.

\begin{figure}
\centering
\includegraphics[width=8cm]{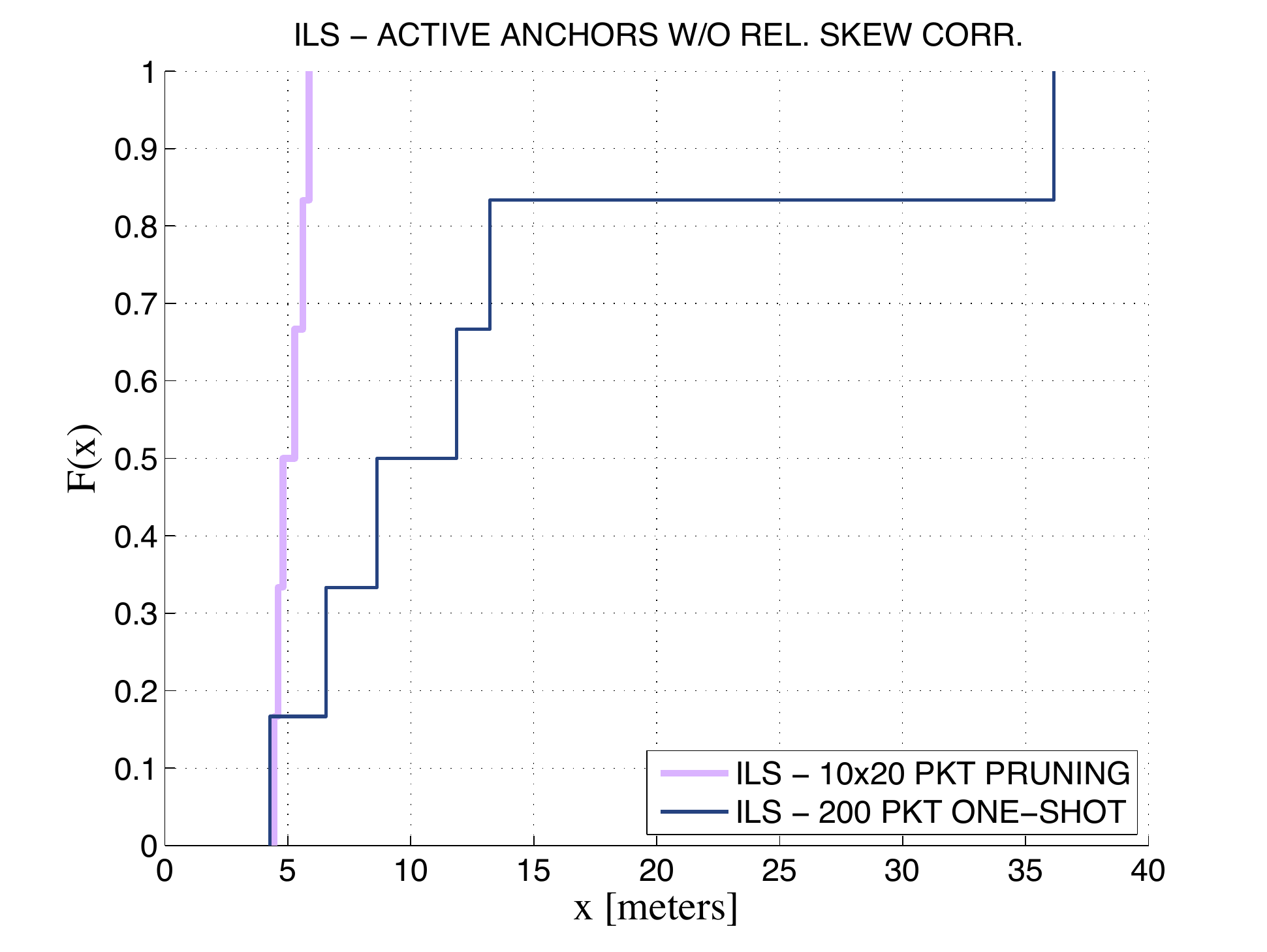}
\caption{Distribution of absolute positioning error obtained in the same conditions of Fig. \ref{fig:activegarden} but without  skew correction.}
\label{fig:nocorr}
\end{figure}

From this proof we can conclude that the DTDOA system is practically {\em insensitive to a small common frequency offset} (with respect to the nominal frequency)  affecting all nodes. 
On the other hand, DTDOA is extremely sensitive to even small {\em relative frequency offsets} between the nodes. 
For this reason, it is important to estimate and compensate for the {\em relative} differences in clock skew terms, that in our method is achieved by preliminary estimation of and rescaling by $\hat{\gamma}_k$ (ref. Section \ref{sec:gamma}). The penalty for skipping this phase --- equivalently, neglecting the presence of independent  clock skew terms in the clock error model  --- is a dramatic increase in the final position error: to illustrate, we report in Fig. \ref{fig:nocorr} the error distribution that is obtained from the GARDEN data, in the same setting of Fig. \ref{fig:activegarden}, when skipping the skew correction phase. Comparing Fig. \ref{fig:activegarden} and Fig. \ref{fig:nocorr} (note the different horizontal scale) we observe a tenfold increase of the median S\&P error, while the worst-case error with one-shot procedure jumps from 1.4 to 36 meters. This confirms the importance of estimating and adjusting for (relative) clock skews.

%
%


\section*{Acknowledgments}
The work by F. Gringoli and N. Facchi was partly supported by the European Union under projects FP7-258301 (CREW) and H2020-645274 (WiSHFUL).
F. Ricciato would like to thank Dr. Reinhard Exel for some fruitful discussions about the impact and mitigation of NLOS in DTDoA scenarios. 

\section*{Additional Resources}
The experimental datasets  and all the software developed to setup the testbed  are available for non-anonymous download from  \url{http://www.ing.unibs.it/~openfwwf/localisation}. 


\bibliographystyle{unsrt}
\bibliography{bibliopassivelocc}

%
%
%
%

\end{document}